\documentclass[aps,prb,preprint,groupedaddress]{revtex4-2}

\usepackage{graphicx}
\usepackage{dcolumn}
\usepackage{bm}

\begin{document}


\title{Fano-Kondo resonance versus Kondo plateau in \\
an Aharonov-Bohm ring with an embedded quantum dot}%

\author{Mikio Eto}
\affiliation{Faculty of Science and Technology, Keio University,
3-14-1 Hiyoshi, Kohoku-ku, Yokohama 223-8522, Japan}
\author{Rui Sakano}
\affiliation{Institute for Solid State Physics, University of Tokyo,
Kashiwa, Chiba 277-8581, Japan}

\date{Received 10 September 2020; accepted 5 November 2020}%

\begin{abstract}
We theoretically examine the transport through an Aharonov-Bohm ring with an
embedded quantum dot (QD), the so-called QD interferometer, to address two
controversial issues regarding the shape of the Coulomb peaks and
measurement of the transmission phase shift through a QD.
We extend a previous model
[B.\ R.\ Bulka and P.\ Stefa\'nski, Phys.\ Rev.\ Lett.\ \textbf{86}, 5128
(2001); W.\ Hofstetter, J.\ K\"{o}nig, and H.\ Schoeller, {\it ibid}.\
\textbf{87}, 156803 (2001)] to consider multiple conduction channels
in two external leads, $L$ and $R$. We introduce a parameter
$p_{\alpha}$ ($|p_{\alpha}| \le 1$) to characterize a connection
between the two arms of the ring through lead $\alpha$ ($=L$, $R$), which
is the overlap integral between the conduction modes coupled to the two arms.
First, we study the shape of a conductance peak as a function of energy level
in the QD, in the absence of electron-electron interaction $U$.
We show an asymmetric Fano resonance for $|p_{L,R}| = 1$ in the case of
single conduction channel in the leads and an almost symmetric Breit-Wigner
resonance for $|p_{L,R}| < 0.5$ in the case of multiple channels.
Second, the Kondo effect is taken into account by the Bethe ansatz exact
solution in the presence of $U$.
We precisely evaluate the conductance at temperature $T=0$ and
show a crossover from an asymmetric Fano-Kondo resonance to
the Kondo plateau with changing $p_{L,R}$.
Our model is also applicable to the
multi-terminal geometry of the QD interferometer. We discuss the measurement
of the transmission phase shift through the QD in a three-terminal geometry by
a ``double-slit experiment.'' We derive an analytical expression for the
relation between the measured value and the intrinsic value of the phase shift.
\end{abstract}

\maketitle

\section{Introduction}

In the mesoscopic physics, an Aharonov-Bohm (AB) ring with an embedded
quantum dot (QD), the so-called QD interferometer, has been intensively
studied to elucidate the coherent transport through a QD with discrete
energy levels and strong Coulomb interaction \cite{PhysRevLett.74.4047,
Nature.385.417,PhysRevLett.88.166801,PhysRevB.66.115311}.
Controversial issues still remain regarding the transport through
the interferometer despite long-term experimental and theoretical studies.
We theoretically revisit these issues by generalizing a previous
model to consider multiple conduction channels in external leads and
a multi-terminal geometry.

We first discuss the shape of Coulomb peaks, i.e.,
conductance $G$ as a function of gate voltage attached to
the QD to control the energy levels electrostatically.
Kobayashi {\it et al}.\ observed an asymmetric shape of the Coulomb
peaks, which has a peak and dip in accordance with the Fano resonance,
using a QD interferometer \cite{PhysRevLett.88.256806}.
The Fano resonance stems from the interference between a discrete energy
level in the QD and continuum energy states in the ring
\cite{PhysRev.124.1866,RevModPhys.82.2257}.
Remarkably the resonant shape of the Coulomb peaks changes
with a magnetic flux penetrating the ring.
However, the other groups observed symmetric Coulomb peaks,
which can be fitted to the Lorentzian function of Breit-Wigner resonance
\cite{PhysRevLett.113.126601}. No criteria has been elucidated regarding
the Fano or Breit-Wigner resonance in the QD interferometer.

The second issue concerns the measurement of the transmission
phase shift through a QD using the QD interferometer
as a double-slit experiment.
It is well known that the phase shift cannot be observed
by the interferometer in the two-terminal geometry \cite{PhysRevLett.74.4047}.
This is due to the restriction by the Onsager's reciprocity theorem:
Conductance $G$ satisfies $G({\bm B})=G(-{\bm B})$ for magnetic field
${\bm B}$, or $G(\phi)=G(-\phi)$ for the AB phase $\phi=2\pi\Phi/(h/e)$ with
magnetic flux $\Phi$ penetrating the ring
\cite{PhysRevLett.88.166801,PhysRevB.66.115311}.
The phase measurement was first reported using the interferometer in a
four-terminal geometry \cite{Nature.385.417}.
In the Kondo regime, the phase shift through the QD should be locked at
$\pi/2$ \cite{PhysRevLett.84.3710,PhysRevLett.90.106602,PhysRevB.80.115330}.
This phase locking was also investigated experimentally
using four- or three-terminal devices \cite{PhysRevLett.113.126601,
Ji779,PhysRevLett.88.076601,PhysRevLett.96.126801,PhysRevLett.100.226601,
doi:10.1063/1.4928035,PhysRevB.94.081303}.
It is nontrivial, however, how precisely the phase shift is measured
using the multi-terminal interferometer.

Theoretically,
Bulka and Stefa\'nski studied Fano and Kondo resonances using
a model for the two-terminal QD interferometer, in which a QD is
coupled to leads $L$ and $R$ and the leads
are directly coupled to each other \cite{PhysRevLett.86.5128}.
Hofstetter {\it et al}.\ found an asymmetric Fano-Kondo resonance
by applying the numerical renormalization group calculation to
an equivalent model \cite{PhysRevLett.87.156803}.
Their works were followed by many theoretical studies,
e.g., to elucidate various aspects of the Kondo effect
\cite{PhysRevB.71.121309,
doi:10.1143/JPSJ.77.123714,PhysRevLett.102.166806,PhysRevB.81.155323,
PhysRevB.81.113402,PhysRevB.81.165107,PhysRevB.82.165426,PhysRevB.83.165310,
PhysRevB.88.245104,PhysRevB.95.075147},
fluctuation theorem \cite{PhysRevB.79.235311},
and dynamics of electronic states \cite{PhysRevB.85.155324}.
Recently, the Fano resonance was proposed to detect
the Majorana bound states \cite{PhysRevB.96.085417,PhysRevB.98.075142}.

\begin{figure}[t]%
\centering
\includegraphics*[width=0.3\linewidth]{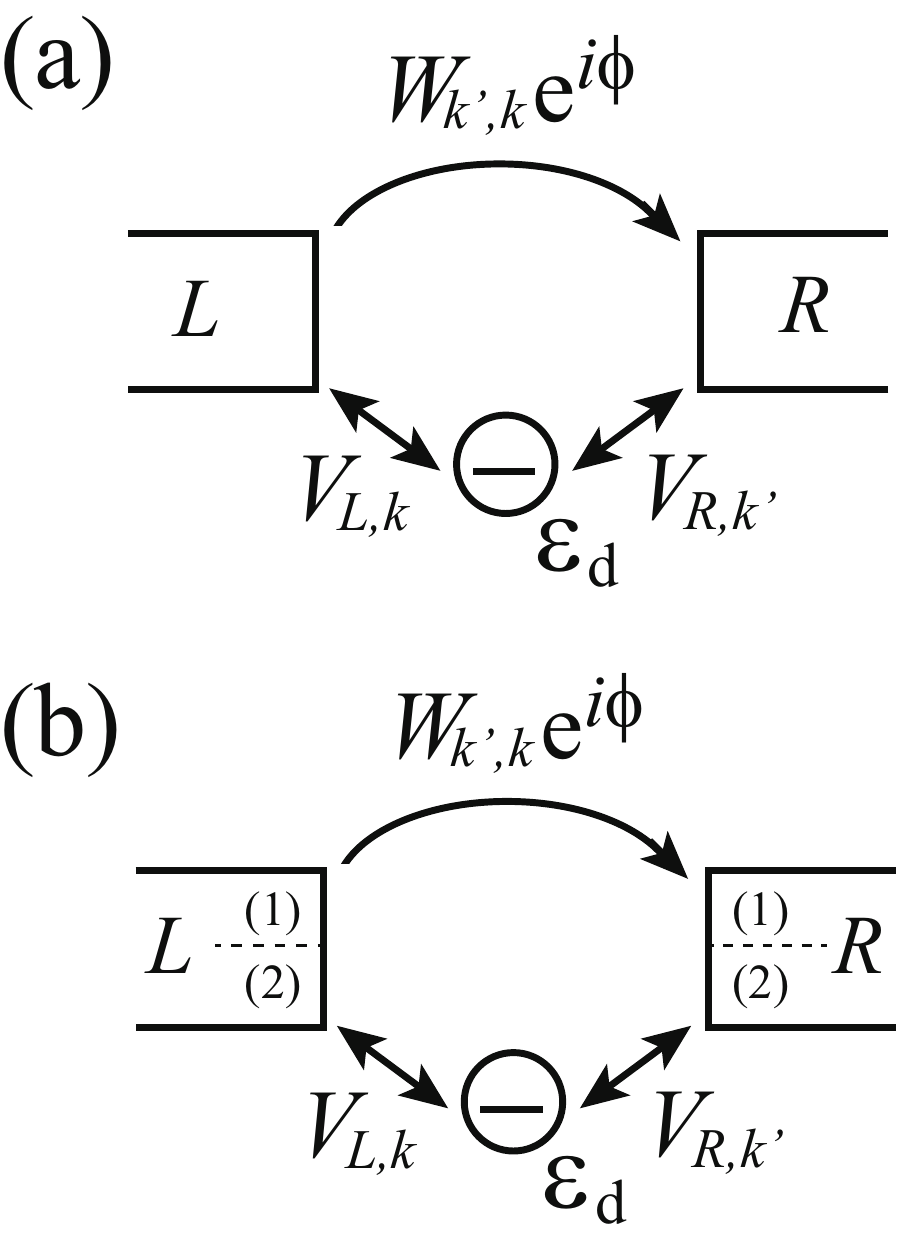}
\caption{%
(a) Model for an AB ring with an embedded quantum dot (QD), the so-called
QD interferometer, in the two-terminal geometry.
The lower arm of the ring involves a QD with single energy level
$\varepsilon_d$, whereas the upper arm directly connects leads $L$ and $R$.
The tunnel couplings between the QD and leads, $V_{L,k}$, $V_{R,k'}$,
and that through the upper arm, $W_{k',k}$, depend on states $k$ in lead $L$
and state $k'$ in lead $R$.
A magnetic flux $\Phi$ penetrating the ring is taken into account by
the AB phase $\phi=2\pi\Phi/(h/e)$. 
The electron-electron interaction $U$ works in the QD.
(b) Model for the QD interferometer in a four-terminal geometry,
with leads $L(1)$, $L(2)$, $R(1)$, and $R(2)$.
State $k$ [$k'$] belongs to lead $L(1)$ or $L(2)$ [$R(1)$ or $R(2)$].
The chemical potentials in the leads are denoted by
$\mu_{L}^{(1)}$, $\mu_{L}^{(2)}$, $\mu_{R}^{(1)}$, and $\mu_{R}^{(2)}$,
respectively, in the formulation in Appendix A.
We fix $\mu_{L}^{(1)}=\mu_{L}^{(2)} \equiv \mu_{L}$ and
$\mu_{R}^{(1)}=\mu_{R}^{(2)} \equiv \mu_{R}$ with $\mu_{L}-\mu_{R}=eV$
in our calculations.
}
\label{Fig1:model0}
\end{figure}

Although the model in Refs.\ \cite{PhysRevLett.86.5128,PhysRevLett.87.156803}
was widely used,
it is insufficient to describe experimental situations with multiple conduction
channels in the leads.
In the present paper, we propose an extended model for the
QD interferometer to resolve the above-mentioned problems.
As shown in Fig.\ \ref{Fig1:model0}(a), our model is the same as the
previous model except the tunnel couplings, $V_L$, $V_R$, and $W$, depend
on the states in leads $L$ and $R$. We show that the state-dependence can be
disregarded only in the case of single conduction channel in the leads.

Our model yields a parameter $p_{\alpha}$ ($|p_{\alpha}| \le 1$) to
characterize a connection between the two arms of the ring through lead
$\alpha$ ($=L$, $R$), which is the overlap integral between the
conduction modes coupled to the upper and lower arms of the ring.
First, we examine the shape of a conductance peak in the two-terminal geometry,
in the absence of electron-electron interaction $U$ in the QD.
We show an asymmetric Fano resonance for $|p_{L,R}| = 1$
in the case of single conduction channel in the leads and
an almost symmetric Breit-Wigner
resonance at $|p_{L,R}| < 0.5$ in the case of multiple channels.
Hence our model could explain the experimental results of both the
asymmetric Fano resonance \cite{PhysRevLett.88.256806} and almost symmetric
Breit-Wigner resonance \cite{PhysRevLett.113.126601}, with fitting parameters
$p_{L,R}$ to their data.

Second, the transport in the Kondo regime is examined by exploiting
the Bethe ansatz exact solution.
This method precisely gives us the conductance at temperature $T=0$ in
the presence of $U$.
We show a crossover from an asymmetric Fano-Kondo resonance
\cite{PhysRevLett.87.156803} to the Kondo plateau with changing $p_{L,R}$.

Our model is also applicable to the multi-terminal geometry,
where state $k$ [$k'$] belongs to lead $L(1)$ or $L(2)$
[$R(1)$ or $R(2)$], as depicted in Fig.\ \ref{Fig1:model0}(b).
We discuss the measurement of the transmission phase shift through
the QD by a ``double-slit experiment'' using a three-terminal interferometer.
We derive an analytical relation between the observed phase shift and
intrinsic phase shift in the absence of $U$.
Using a simple model to represent the experiment by Takada
{\it et al}.\ \cite{PhysRevLett.113.126601,doi:10.1063/1.4928035,
PhysRevB.94.081303}, we evaluate the measured phase shift
in both the absence and presence of $U$. For $U \ne 0$, we show
that the phase locking at $\pi/2$ is observable in the Kondo regime
although it is slightly different from the behavior of the intrinsic
phase shift that satisfies the Friedel sum rule.

The organization of the present paper is as follows. In section II,
we present our model and calculation method.
The parameters $p_{L}$ and $p_R$ are introduced, which are relevant
to the shape of a conductance peak. We explain the calculation method of the
conductance at $T=0$, taking into account the Kondo effect exactly.
In section III, the calculated results are given for the shape of the
conductance peak in a two-terminal geometry. We discuss the asymmetric Fano
resonance versus symmetric Breit-Wigner resonance in the absence of $U$,
by changing $p_{L,R}$. We also study the conductance in the
Kondo regime in the presence of $U$ and show a crossover from
an asymmetric Fano-Kondo resonance to the Kondo plateau.
In section IV, we examine the phase measurement in a three-terminal
geometry by a double-slit interference experiment.
We derive an analytical relation between the measured value and
intrinsic value for the transmission phase shift through the QD
in the absence of $U$. Two specific models are studied to see a
crossover from two- to three-terminal measurement and to simulate
the experimental situation using two quantum wires to form the
QD interferometer \cite{PhysRevLett.113.126601,doi:10.1063/1.4928035,
PhysRevB.94.081303}.
Section V is devoted to the discussion regarding the justification and
generality of our model. The conclusions are given in section VI.

\section{Model and calculation method}
\subsection{Model}

Let us consider a model for the QD interferometer in a two-terminal geometry,
depicted in Fig.\ \ref{Fig1:model0}(a). The Hamiltonian is given by
\begin{equation}
H =
H_{\rm dot}+H_{\rm leads}+H_{\rm T},
\label{eq:Hamiltonian}
\end{equation}
where
\begin{eqnarray}
H_{\rm dot} &=& \varepsilon_d \sum_{\sigma} n_{\sigma}
+ U n_{\uparrow} n_{\downarrow},
\\
H_{\rm leads} &=& 
  \sum_{\alpha=L,R} \sum_{k \sigma}
  \varepsilon_k a_{\alpha,k\sigma}^{\dagger}a_{\alpha,k\sigma}
\\
H_{\rm T} &=& \sum_{\alpha=L,R} \sum_{k \sigma}
    (V_{\alpha,k} a_{\alpha,k \sigma}^{\dagger} d_{\sigma}+{\rm H.c.})
  + \sum_{k,k',\sigma} (W_{k',k}e^{i\phi}
    a_{R,k' \sigma}^{\dagger}a_{L,k \sigma}+{\rm H.c.})
\label{eq:tunnelH}
\end{eqnarray}
Here, 
$n_{\sigma}=d_{\sigma}^{\dagger}d_{\sigma}$ is the number operator
for electrons with spin $\sigma$ in the QD with energy level
$\varepsilon_d$, where $d_{\sigma}^{\dagger}$ and
$d_{\sigma}$ are creation and annihilation operators, respectively.
$a_{\alpha,k\sigma}^{\dagger}$ and $a_{\alpha,k\sigma}$ are those
for conduction electrons in lead $\alpha$ ($=L$, $R$) with state $k$
and spin $\sigma$, whose energy is denoted by $\varepsilon_k$.
$U$ is the electron-electron interaction in the QD.
The tunnel Hamiltonian, $H_{\rm T}$, connects the QD and
state $k$ in lead $\alpha$ by $V_{\alpha,k}$ through the lower arm
of the ring, whereas it connects
state $k$ in lead $L$ and state $k'$ in lead $R$ by $W_{k',k}$
through the upper arm of the ring.
The AB phase is defined by $\phi=2\pi\Phi/(h/e)$ for a magnetic flux
$\Phi$ penetrating the ring.
To make the calculation simple, we decompose $W_{k',k}$ into
the contributions from state $k$ in lead $L$ and state $k'$
in lead $R$ as
\begin{equation}
W_{k',k}=\sqrt{w_{R,k'}w_{L,k}}.
\label{eq:Wseparable}
\end{equation}
This separable form is justified for the tight-binding models, as
discussed in section V.

For lead $\alpha$, we introduce the following three parameters to
describe the contribution to the transport:
\begin{eqnarray}
\Gamma_{\alpha}(\varepsilon) &=& 
\pi \sum_k (V_{\alpha,k})^2 \delta(\varepsilon-\varepsilon_k),
\label{eq:Gamma}
\\
x_{\alpha}(\varepsilon) &=& 
\pi \sum_k w_{\alpha,k} \delta(\varepsilon-\varepsilon_k),
\label{eq:introduce_x}
\\
\sqrt{\Gamma_{\alpha}(\varepsilon) x_{\alpha}(\varepsilon)}
p_{\alpha}(\varepsilon) &=& 
\pi \sum_k
V_{\alpha,k} \sqrt{w_{\alpha,k}} \delta(\varepsilon-\varepsilon_k).
\label{eq:introduce_p}
\end{eqnarray}
We assume that the $\varepsilon$-dependence of these parameters is weak
around the Fermi level and simply express $\Gamma_{\alpha}$, $x_{\alpha}$,
and $p_{\alpha}$ for $\varepsilon \approx E_{\rm F}$.
$\Gamma_{\alpha}$ ($x_{\alpha}$) characterizes the strength of tunnel
coupling to the QD (coupling through the upper arm of the ring).
Using $x=x_L x_R$,
the transmission probability through the upper arm of the ring is given by
\begin{equation}
T_{\rm upper}=\frac{4x}{(1+x)^2}.
\label{eq:T_upper}
\end{equation}
Concerning $x_L$ and $x_R$, the physical quantities
are always written in terms of $x=x_L x_R$  in our model \footnote{In Eq.\
(\ref{eq:Wseparable}), $w_{R,k'}$ and $w_{L,k}$ can be replaced by
$\lambda w_{R,k'}$ and $w_{L,k}/\lambda$ with a constant $\lambda$ for a given
$W_{k',k}$. As a result, $x_L$ and $x_R$ are not well-defined by themselves
in Eq.\ (\ref{eq:introduce_x}) in contrast to $p_L$ and $p_R$ in
Eq.\ (\ref{eq:introduce_p}). $x_L$ and $x_R$ always appear in the form of
$x=x_L x_R$ in the physical quantities.}.

The parameter $p_{\alpha}$ ($|p_{\alpha}| \le 1$) defined by Eq.\
(\ref{eq:introduce_p}) characterizes a connection
between the two arms of the ring through lead $\alpha$ ($=L$, $R$).
Namely, $p_{\alpha}(\varepsilon)$ is an overlap integral between
the conduction mode coupled to the QD and that coupled to the upper arm of
the ring in lead $\alpha$ at a given energy $\varepsilon$.
The tunnel Hamiltonian $H_{\rm T}$ in Eq.\ (\ref{eq:tunnelH}) indicates
that these modes are given by
$|\psi_{\alpha}^{\rm QD}(\varepsilon) \rangle \propto
\sum_{k} V_{\alpha,k}| \alpha,k \rangle \delta(\varepsilon-\varepsilon_k)$
and $|\psi_{\alpha}^{\rm upper}(\varepsilon) \rangle \propto
\sum_{k} \sqrt{w_{\alpha,k}} | \alpha,k \rangle
\delta(\varepsilon-\varepsilon_k)$, respectively, where
$| \alpha,k \rangle$ is the state $k$ in lead $\alpha$.
For $|\psi(\varepsilon) \rangle=\sum_{k} C_k | \alpha,k \rangle
\delta(\varepsilon-\varepsilon_k)$ and
$|\varphi(\varepsilon) \rangle=\sum_{k} D_k | \alpha,k \rangle
\delta(\varepsilon-\varepsilon_k)$, we denote the inner product by
$\langle \psi(\varepsilon) | \varphi(\varepsilon') \rangle
=\langle \psi | \varphi \rangle_{\varepsilon}
\delta(\varepsilon-\varepsilon')$, or
$\langle \psi | \varphi \rangle_{\varepsilon}
=\sum_{k} C_k^* D_k \delta(\varepsilon-\varepsilon_k)$. Then
\begin{equation}
p_{\alpha}(\varepsilon)=
\frac{\langle \psi_{\alpha}^{\rm QD} | \psi_{\alpha}^{\rm upper}
\rangle_{\varepsilon}}
{\sqrt{\langle \psi_{\alpha}^{\rm QD} | \psi_{\alpha}^{\rm QD}
\rangle_{\varepsilon}
\langle \psi_{\alpha}^{\rm upper} | \psi_{\alpha}^{\rm upper}
\rangle_{\varepsilon}}}.
\label{eq:p_overlap_integral}
\end{equation}
The interference by the AB effect is maximal when $|p_L|=|p_R|=1$,
whereas it completely disappears when $p_{L}=0$ or $p_{R}=0$.
In the previous model \cite{PhysRevLett.86.5128,PhysRevLett.87.156803},
$|\psi_{\alpha}^{\rm QD} \rangle =|\psi_{\alpha}^{\rm upper} \rangle$
and thus $p_{\alpha}=1$ since $V_{\alpha,k}$ and $\sqrt{w_{\alpha,k}}$ are
constant, irrespective of state $k$. As seen in the following sections,
$p_L$ and $p_R$ play a crucial role in determining the shape of
conductance peaks.
Although $p_L$ and $p_R$ should be given by the details of experimental
systems, we treat them as parameters as well as $\Gamma_L$, $\Gamma_R$,
and $x$.

As an example, let us consider quasi-one-dimensional leads, or leads of
a quantum wire. The state in lead $\alpha$ is specified by $k=q$
in the case of single conduction channel and by $k=(q,i)$ in the presence
of multiple channels, where $q$ is the momentum along the wire and
$i$ is the index of the subbands. In the former,
$V_{\alpha,k}=V_{\alpha}(\varepsilon_k)$ and
$w_{\alpha,k}=w_{\alpha}(\varepsilon_k)$, which yield
$\Gamma_{\alpha}(\varepsilon)=\pi \rho_{\alpha}(\varepsilon)
[V_{\alpha}(\varepsilon)]^2$ with density of states $\rho_{\alpha}$
in the lead, $x_{\alpha}(\varepsilon)=
\pi \rho_{\alpha}(\varepsilon) w_{\alpha}(\varepsilon)$,
and $|p_{\alpha}|=1$ from Eqs.\ (\ref{eq:Gamma})--(\ref{eq:introduce_p}).
In the case of multiple channels, $|p_{\alpha}|<1$, as shown in section V.
Note that a similar parameter to $p_{\alpha}$ was introduced
in the study on a double quantum dot in parallel and was evaluated for
three- or two-dimensional leads with a flat surface \cite{PhysRevB.74.205310}.

For the multi-terminal geometry,
lead $\alpha$ is divided into leads $\alpha(1)$
and $\alpha(2)$, as depicted in Fig.\ \ref{Fig1:model0}(b).
The Hamiltonian in Eq.\
(\ref{eq:Hamiltonian}) is applicable even to this case, in which
the summation over $k$ is taken in both lead $\alpha(1)$ 
[denoted by $\sum_k^{(1)}$] and lead $\alpha(2)$
[by $\sum_k^{(2)}$].
We define $\Gamma_{\alpha}^{(j)}$ using Eq.\ (\ref{eq:Gamma}) with
the summation over $k$ in lead $\alpha(j)$ only:
\begin{equation}
\Gamma_{\alpha}^{(j)}(\varepsilon) =
\pi \sum_k^{(j)} |V_{\alpha,k}|^2 \delta(\varepsilon-\varepsilon_k)
\end{equation}
for $\alpha=L$, $R$ and $j=1$, 2.
Similarly, we define $x_{\alpha}^{(1)}$,
$x_{\alpha}^{(2)}$, $p_{\alpha}^{(1)}$, and $p_{\alpha}^{(2)}$.
They satisfy the following relations.
\begin{eqnarray}
\Gamma_{\alpha} &=& \Gamma_{\alpha}^{(1)} + \Gamma_{\alpha}^{(2)},
\\
x_{\alpha} &=& x_{\alpha}^{(1)} + x_{\alpha}^{(2)},
\\
\sqrt{\Gamma_{\alpha} x_{\alpha}} p_{\alpha}
&=&
\sqrt{\Gamma_{\alpha}^{(1)} x_{\alpha}^{(1)}} p_{\alpha}^{(1)}
 +\sqrt{\Gamma_{\alpha}^{(2)} x_{\alpha}^{(2)}} p_{\alpha}^{(2)}.
\end{eqnarray}

\subsection{Formulation of electric current}

We formulate the electric current using the Keldysh Green's functions
along the lines of Ref.\ \cite{PhysRevLett.87.156803} (see Appendix A).
For example, the current from lead $L(1)$ in Fig.\ \ref{Fig1:model0}(b) is
given by
\begin{equation}
I_L^{(1)} = -e \langle \dot{N}_L^{(1)} \rangle
=-\frac{e}{i\hbar} \langle [N_L^{(1)}, H] \rangle,
\label{eq:current_L^(1)}
\end{equation}
where
\begin{equation}
N_L^{(1)}=\sum_{k \sigma}^{(1)}
a_{L,k\sigma}^{\dagger}a_{L,k\sigma}
\end{equation}
is the number operator for electrons in the lead. In the stationary state,
$I_L^{(1)}$ is expressed
in terms of the retarded Green's function $G_{d,d}^{\rm r}(\varepsilon)$ and
lesser Green's function $G_{d,d}^{<}(\varepsilon)$ of the QD,
in Eq.\ (\ref{eq:I_L_(1)_final}) in Appendix A.

Next, we eliminate $G_{d,d}^{<}(\varepsilon)$ from the expression
and write the current using $G_{d,d}^{\rm r}(\varepsilon)$ only.
We restrict ourselves to the case of
\begin{equation}
\mu_{L}^{(1)}=\mu_{L}^{(2)} \equiv \mu_{L}, \ \
\mu_{R}^{(1)}=\mu_{R}^{(2)} \equiv \mu_{R},
\label{eq:mu_L_mu_R}
\end{equation}
with $\mu_{L}-\mu_{R}=eV$, to simplify the current expression.
Then the current conservation is written
as follows in the stationary state:
\begin{eqnarray}
0 &=& I_L^{(1)}+I_L^{(2)}+I_R^{(1)}+I_R^{(2)}
\nonumber \\
&=& \frac{4e}{h} \int d\varepsilon
\Biggl\{
-\tilde{\Gamma}
\Bigl[ \left[ f_L(\varepsilon)+f_R(\varepsilon) \right]
{\rm Im} G_{d,d}^{\rm r}(\varepsilon) +
{\rm Im} G_{d,d}^{<}(\varepsilon) \Bigr]
\nonumber \\
& & +
\left[ f_L(\varepsilon)-f_R(\varepsilon) \right]
\left[ -(\Gamma_L-\Gamma_R)
-4\frac{\sqrt{\Gamma_{L} \Gamma_{R}x} p_L p_R}{(1+x)^2} \sin \phi
+\frac{x(x+3)}{(1+x)^2}
 \left( \Gamma_{L}p_L^2 - \Gamma_{R}p_R^2 \right)
\right]
\nonumber \\
& & \times {\rm Im} G_{d,d}^{\rm r}(\varepsilon)
 \Bigg\},
\label{eq:current_conserve}
\end{eqnarray}
where $f_{\alpha}(\varepsilon)=
[(\varepsilon-\mu_{\alpha})/(k_{\rm B}T)+1]^{-1}$
is the Fermi distribution function in lead $\alpha(1)$ or $\alpha(2)$
[$\tilde{\Gamma}$ will be given in Eq.\ (\ref{eq:effective_Gamma})].
Using Eq.\ (\ref{eq:current_conserve}), we eliminate
$G_{d,d}^{<}(\varepsilon)$ from the current expression,
e.g., Eq.\ (\ref{eq:I_L_(1)_final}) for $I_L^{(1)}$.

\subsection{Electric current in two-terminal systems}

Now we present the expression for the electric current in the two-terminal
systems. The current from the left lead
$I_L$ [$=I_L^{(1)}+I_L^{(2)}$ using the results in Appendix A] is
expressed as
\begin{equation}
I_L = \frac{2e}{h} \int d\varepsilon
\left[ f_L(\varepsilon)-f_R(\varepsilon) \right]
T(\varepsilon),
\end{equation}
with the transmission probability
\begin{equation}
T(\varepsilon) =
\frac{4x}{(1+x)^2}+8\frac{1-x}{(1+x)^3}
\sqrt{\Gamma_{L} \Gamma_{R}x} p_L p_R \cos \phi
{\rm Re} G_{d,d}^{\rm r}(\varepsilon)
+\frac{4 C_1}{(1+x)^3\tilde{\Gamma}}
{\rm Im} G_{d,d}^{\rm r}(\varepsilon).
\label{eq:2terminal_IL}
\end{equation}
Here, the coefficient $C_1$ is given by
\begin{eqnarray}
C_1 &=&
\frac{x^3}{1+x}
\left[ (\Gamma_{L} p_L^2)^2 + (\Gamma_{R} p_R^2)^2 \right]
+x(1-x) \left[ (\Gamma_{L} p_L)^2 + (\Gamma_{R} p_R)^2 \right]
\nonumber \\
& & 
-\Gamma_{L} \Gamma_{R} \Bigg[
(1+x)^3  +\frac{4x}{1+x} (p_L p_R)^2 \sin^2 \phi
+\frac{x^2(x^2+4x+9)}{1+x} (p_L p_R)^2
\nonumber \\
& &
\hspace*{1.8cm}
-x(x^2+3x+4) (p_L^2 + p_R^2) \Biggr].
\end{eqnarray}
Note that (i) for $p_L=p_R=1$, where a single conduction channel is
effective in each lead, Eq.\ (\ref{eq:2terminal_IL})
coincides with the current expression derived in Ref.\
\cite{PhysRevLett.87.156803}.
(ii) For $p_L=p_R=0$, the transmission probability is given by
\begin{eqnarray}
T(\varepsilon)=
\frac{4x}{(1+x)^2}-\frac{4\Gamma_{L} \Gamma_{R}}{\Gamma_{L} +\Gamma_{R}}
{\rm Im} G_{d,d}^{\rm r}(\varepsilon).
\label{eq:T_p_L=p_R=0}
\end{eqnarray}
This is the summation of the
transmission probability through the upper arm,
$T_{\rm upper}$ in Eq.\ (\ref{eq:T_upper}), and that through the QD,
indicating no interference effect between the two paths in the QD
interferometer.

For multi-terminal systems, the current is expressed in terms of
the retarded Green's function $G_{d,d}^{\rm r}(\varepsilon)$ in a similar way.
The expression is given in Eqs.\ (\ref{eq:3terminal_IRa}) and
(\ref{eq:3terminal_IRb}) in Appendix A.

\subsection{Exact calculation for Kondo effect}

In the absence of Coulomb interaction, $U=0$, the retarded Green's function
of the QD is given by
$G_{d,d}^{\rm r}(\varepsilon)=1/(\varepsilon-\varepsilon_d-\Sigma_d)$,
where the self-energy by the tunnel couplings is
\begin{equation}
\Sigma_d=-\frac{2 \sqrt{\Gamma_L \Gamma_R x}}{1+x}
p_L p_R \cos\phi-i\tilde{\Gamma},
\label{eq:self_energy_G_dd}
\end{equation}
with an effective linewidth
\begin{equation}
\tilde{\Gamma} =
  \Gamma_L \left(1- \frac{x}{1+x} p_L^2 \right)
  +\Gamma_R \left(1- \frac{x}{1+x} p_R^2 \right).
\label{eq:effective_Gamma}
\end{equation}
This expression is common to two- and multi-terminal systems.

In the presence of $U$, $G_{d,d}^{\rm r}(\varepsilon)$ is evaluated exactly
in the following way. The Green's function at $U=0$ indicates that our
models are equivalent to the situation in which a QD with an
energy level
\begin{equation}
\tilde{\varepsilon}_d(\phi) =\varepsilon_d
-\frac{2 \sqrt{\Gamma_L \Gamma_R x}}{1+x} p_L p_R \cos\phi
\label{eq:effective_e_d}
\end{equation}
is connected to a lead with linewidth $\tilde{\Gamma}$,
as shown in Appendix B.
In the Fermi liquid theory, the Green's function is written as
\begin{equation}
G_{d,d}^{\rm r}(0)=
\frac{z}{-\tilde{\varepsilon}_d^*+iz\tilde{\Gamma}}
=\frac{\tilde{\Gamma}^*}{\tilde{\Gamma}}
\frac{1}{-\tilde{\varepsilon}_d^*+i\tilde{\Gamma}^*},
\label{eq:Greenfunction_with_U}
\end{equation}
at $\varepsilon=E_{\rm F}=0$,
where $\tilde{\varepsilon}_d^*$ is the renormalized value of
$\tilde{\varepsilon}_d(\phi)$ in Eq.\ (\ref{eq:effective_e_d}),
$\tilde{\Gamma}^*=z\tilde{\Gamma}$ is that of $\tilde{\Gamma}$
in Eq.\ (\ref{eq:effective_Gamma}), and $z$ is a factor of
wavefunction renormalization by the electron-electron interaction $U$
\cite{hewson_1993,PhysRevLett.70.4007,Hewson_2001}.
Since the phase shift $\theta_{\rm QD}$ at the QD is given
by $\tan \theta_{\rm QD}=\tilde{\Gamma}^*/\tilde{\varepsilon}_d^*$,
the Green's function is determined by $\theta_{\rm QD}$ as
\begin{equation}
G_{d,d}^{\rm r}(0)=
\frac{-1}{\tilde{\Gamma}}e^{i\theta_{\rm QD}} \sin \theta_{\rm QD}.
\label{eq:Greenfunction_sum_rule}
\end{equation}
$\theta_{\rm QD}$ is related to the electron number per spin in the QD
through the Friedel sum rule, $\theta_{\rm QD}=\pi \langle n_{\sigma} \rangle$.
$\langle n_{\sigma} \rangle$ is evaluated at temperature $T=0$ using
the Bethe ansatz exact solution \cite{doi:10.1143/JPSJ.52.1119,Wiegmann_1983}.
Hence we can precisely calculate $G_{d,d}^{\rm r}(0)$ and thus
the conductance $G=dI_L/dV$ ($V \rightarrow 0$) at $T=0$
in the presence of $U$.

It is worth mentioning that
the effective energy level $\tilde{\varepsilon}_d(\phi)$
in the QD gives rise to the $\phi$-dependent Kondo temperature
\cite{PhysRevB.71.121309}. It is written as
\begin{equation}
k_{\rm B} T_{\rm K}(\phi) \approx D \left[
\frac{\tilde{\Gamma}U}
{|\tilde{\varepsilon}_d(\phi)|[\tilde{\varepsilon}_d(\phi)+U]}
\right]^{1/2}
e^{-\pi|\tilde{\varepsilon}_d(\phi)|[\tilde{\varepsilon}_d(\phi)+U]/
(2\tilde{\Gamma}U)}
\end{equation}
with $D$ being the bandwidth \cite{hewson_1993,Haldane_1978}
although $T_{\rm K}(\phi)$ is irrelevant to our study on the transport
properties at $T=0$.

\begin{figure}[t]%
\centering
\includegraphics*[width=0.4\linewidth]{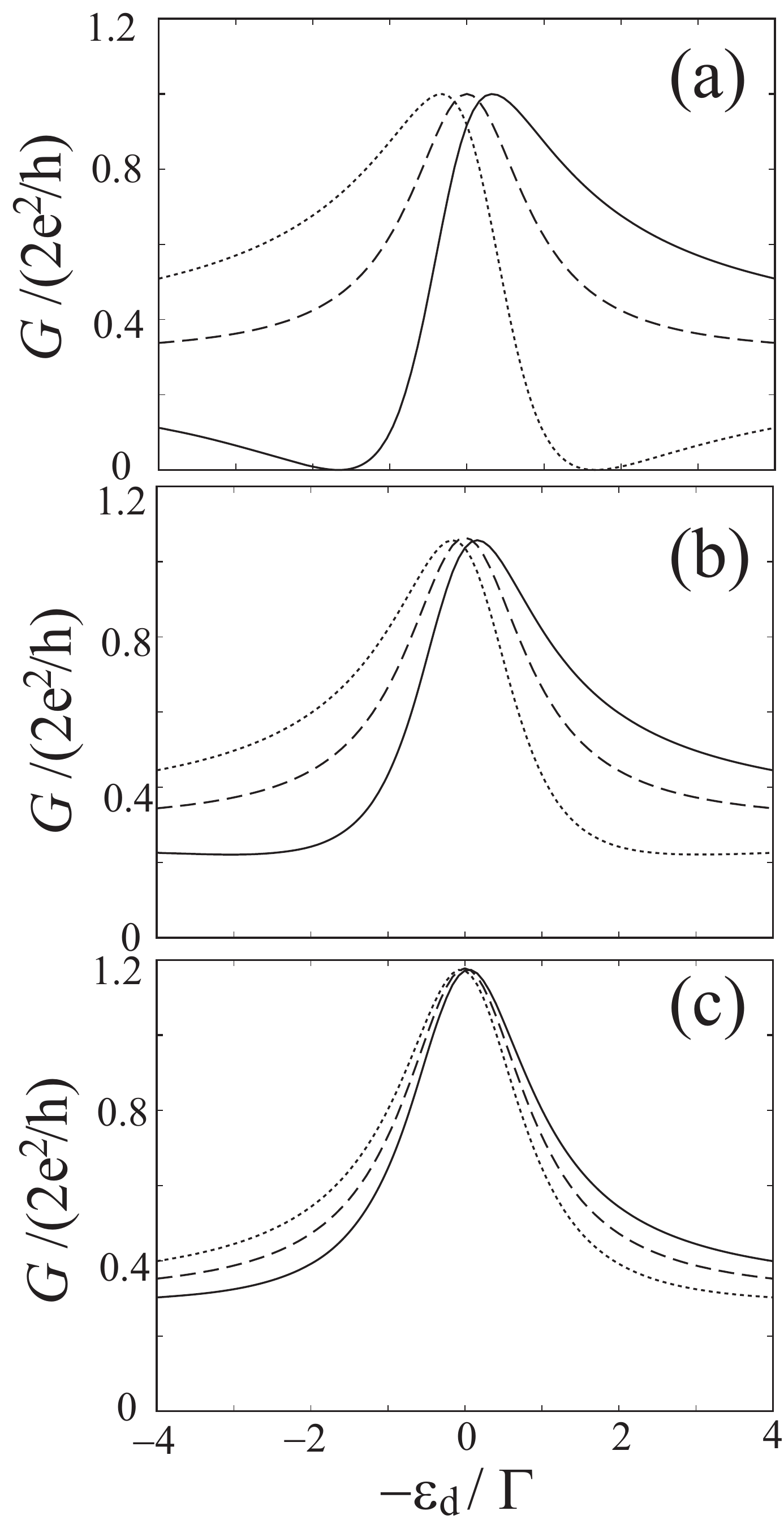}
\caption{%
Calculated results for the conductance $G$ in the two-terminal system
in the absence of $U$. $G$ at temperature $T=0$
is plotted as a function of energy level $\varepsilon_d$ in the
quantum dot. $\Gamma_L=\Gamma_R=\Gamma/2$, $x=0.09$
($x_L=x_R=0.3$), and (a) $p_L=p_R=1$, (b) 0.75, and (c) 0.5.
The AB phase for the magnetic flux penetrating the ring is
$\phi=0$ (solid line), $\phi=\pm \pi/2$ (broken line), and
$\phi=\pi$ (dotted line).
}
\label{Fig2:2term_UO}
\end{figure}

\section{Calculated results in two-terminal geometry}

In this section, we present the calculated results for the two-terminal system,
paying attention to the shape of a conductance peak as a function of 
energy level $\varepsilon_d$ in the QD. We find that parameters
$p_{L}$ and $p_{R}$ are relevant in both the cases of $U=0$ and $U \ne 0$.

\subsection{Fano versus Breit-Wigner resonance}

We begin with the case of no electron-electron interaction in the QD, $U=0$.
Figure \ref{Fig2:2term_UO} shows the conductance $G$ at $T=0$
as a function of energy level $\varepsilon_d$ in the QD
for (a) $p_L=p_R=1$, (b) 0.75, and (c) 0.5.
The AB phase is $\phi=0$ (solid line), $\pm \pi/2$ (broken line), and
$\pi$ (dotted line).
$G(\phi)=G(-\phi)$ holds by the Onsager's reciprocal theorem.

In panel (a) with $p_L=p_R=1$, the conductance $G$ shows an asymmetric resonant
shape with dip and peak in the absence of magnetic field ($\phi=0$). This is
known as the Fano resonance which is ascribable to the interference between the
tunneling through a discrete level and that through continuous states
\cite{PhysRev.124.1866,RevModPhys.82.2257}.
A magnetic field changes the resonant shape to be symmetric
at $\phi=\pm \pi/2$ and asymmetric with peak and dip at $\phi=\pi$. This Fano
resonance is characterized by a complex Fano factor
\cite{PhysRevLett.88.256806}. Indeed the conductance can be analytically
expressed \cite{Ueda2003} in the form of
\begin{equation}
G=\frac{2e^2}{h}\frac{4x}{(1+x)^2}\frac{|e+q|^2}{e^2+1}
\end{equation}
with $e=[\varepsilon_d-\tilde{\varepsilon}_d(\phi)]/\tilde{\Gamma}$,
where $\tilde{\varepsilon}_d(\phi)=\varepsilon_d
-2 \sqrt{\Gamma_L \Gamma_R x} \cos\phi/(1+x)$ and
$\tilde{\Gamma}=(\Gamma_L+\Gamma_R)/(1+x)$
[Eqs.\ (\ref{eq:effective_e_d}) and (\ref{eq:effective_Gamma})
for $p_L=p_R=1$]. The complex Fano factor is given by
\begin{equation}
q=\frac{\sqrt{\Gamma_L \Gamma_R}}{\tilde{\Gamma}\sqrt{x}}
  \left( \frac{1-x}{1+x}\cos\phi - i \sin\phi \right).
\end{equation}

With a decrease in $p_L$ and $p_R$, the conductance peak becomes more
symmetric and its $\phi$-dependence is less prominent, as shown in
panels (b) and (c). The shape of conductance peak is closer to that of
the Lorentzian function of Breit-Wigner resonance
as $p_L$ and $p_R$ go to zero.

Note that the conductance $G$ can exceed unity in units of
$2e^2/h$ when $p_L$, $p_R<1$, reflecting the multiple conduction channels
in the leads. See Eq.\ (\ref{eq:T_p_L=p_R=0})
in the limit of $p_L=p_R=0$: The upper limit
of $G/(2e^2/h)$ is the sum of the transmission probability through the
QD (unity if $\Gamma_L=\Gamma_R$) and that through the upper
arm, $T_{\rm upper}$ in Eq.\ (\ref{eq:T_upper}).

\begin{figure}[t]%
\centering
\includegraphics*[width=0.4\linewidth]{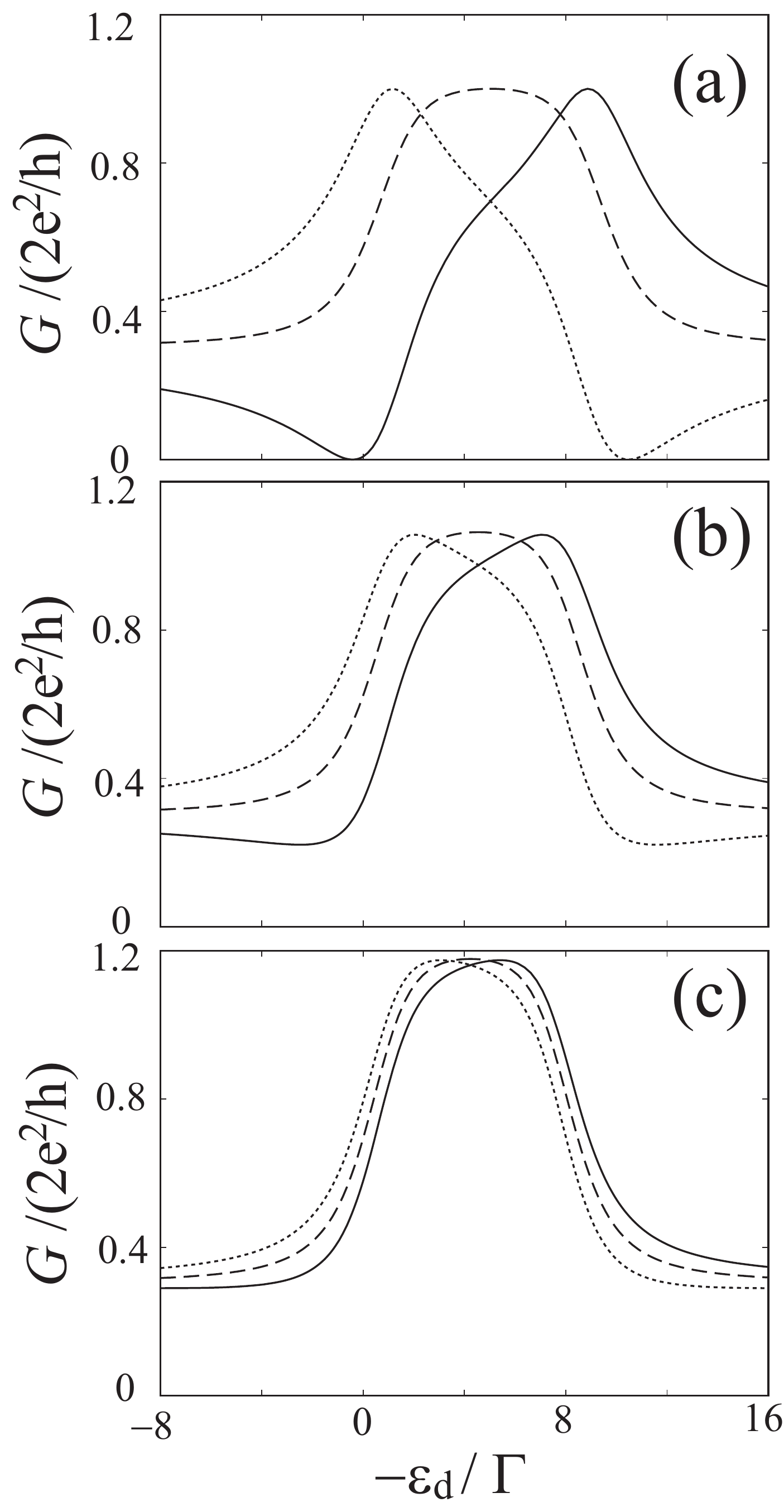}
\caption{%
Calculated results for the conductance $G$ in the two-terminal system
in the presence of $U$.
$G$ at temperature $T=0$ is plotted as a function of energy level
$\varepsilon_d$ in the quantum dot.
$\Gamma_L=\Gamma_R=\Gamma/2$, $x=0.09$
($x_L=x_R=0.3$), and (a) $p_L=p_R=1$, (b) 0.75, and (c) 0.5. $U=8\Gamma$.
The AB phase for the magnetic flux penetrating the ring is
$\phi=0$ (solid line), $\phi=\pm \pi/2$ (broken line), and
$\phi=\pi$ (dotted line).
}
\label{Fig3:2term_U}
\end{figure}

\subsection{Fano-Kondo resonance versus Kondo plateau}

In the presence of $U$, the Kondo effect is exactly taken into account
in the evaluation of the conductance at $T=0$,
as described in the previous section. In Fig.\ \ref{Fig3:2term_U},
the conductance $G$ is shown as a function of energy level $\varepsilon_d$
in the QD, for $U/\Gamma=8$ and $\Gamma_L=\Gamma_R=\Gamma/2$.
(a) $p_L=p_R=1$, (b) 0.75, and (c) 0.5.
The AB phase is $\phi=0$ (solid line), $\pm \pi/2$ (broken line), and
$\pi$ (dotted line).

For $p_L=p_R=1$, $G$ behaves as a ``Fano-Kondo resonance''
proposed by Hofstetter {\it et al}.\ \cite{PhysRevLett.87.156803},
which stems
from an interplay between the Kondo resonance ($G \sim 2e^2/h$ at
$-U < \varepsilon_d < 0$) and the Fano resonance.
When $\phi=0$ ($\pi$), $G$ shows
a dip and peak (peak and dip) with a gradual slope around the center
of the Kondo valley, i.e., Coulomb blockade regime with a spin $1/2$
in the QD. When $\phi=\pi/2$, $G$ is almost constant at
$2e^2/h$ in the Kondo valley and symmetric with respect to the
valley center.

With decreasing $p_L$ and $p_R$, the asymmetric shape of the
Fano-Kondo resonance changes to a conductance plateau,
the so-called Kondo plateau, in the Kondo valley:
$G \rightarrow 2e^2/h + T_{\rm upper}$ as $p_{L,R} \rightarrow 0$ when
$\Gamma_L=\Gamma_R$. Besides, $G$ is less dependent on $\phi$.

\section{Calculated results in three-terminal geometry}

In this section, we examine a three-terminal system
to discuss the measurement of transmission phase shift through the QD
by a ``double-slit interference experiment.''
We assume two leads $R(1)$ and $R(2)$ on the right side and
a single lead $L$ on the left side in Fig.\ \ref{Fig1:model0}(b).
We evaluate the conductance from lead $L$ to $R(1)$ or to $R(2)$,
\begin{equation}
G^{(1)} = -\frac{d I_R^{(1)}}{dV}, \ \
G^{(2)} = -\frac{d I_R^{(2)}}{dV},
\end{equation}
for $eV=\mu_L-\mu_R \rightarrow 0$ ($\mu_{R}^{(1)}=\mu_{R}^{(2)}=\mu_R$)
at $T=0$, as a function of AB phase $\phi$.
We define the measured phase shift by the AB phase $\phi_{\rm max}$
at which the conductance $G^{(1)}(\phi)$ is maximal.

As an intrinsic transmission phase shift through the QD, we introduce
$\theta_{\rm QD}^{(0)}$ and $\theta_{\rm QD}$ by
\begin{eqnarray}
\tan \theta_{\rm QD}^{(0)} &=&
\frac{\Gamma_L + \Gamma_R}{\varepsilon_d-E_{\rm F}},
\label{eq:def_theta_QD(0)}
\\
\tan \theta_{\rm QD} &=&
\frac{\tilde{\Gamma}}{\tilde{\varepsilon}_d(\phi)-E_{\rm F}},
\label{eq:def_theta_QD}
\end{eqnarray}
respectively, in the absence of $U$.
$\theta_{\rm QD}^{(0)}$ is the phase shift through the QD without
the upper arm of the ring, whereas $\theta_{\rm QD}$ satisfies the
Friedel sum rule $\theta_{\rm QD}=\pi \langle n_{\sigma} \rangle$ for the QD
embedded in the ring. The latter depends on the AB phase $\phi$
for the magnetic flux penetrating the ring.
In the next subsection, we derive an analytical relation between the
measured phase $\phi_{\rm max}$ and $\theta_{\rm QD}^{(0)}$ in Eq.\
(\ref{eq:def_theta_QD(0)}) in the absence of $U$.

\begin{figure}[t]%
\centering
\includegraphics*[width=0.4\linewidth]{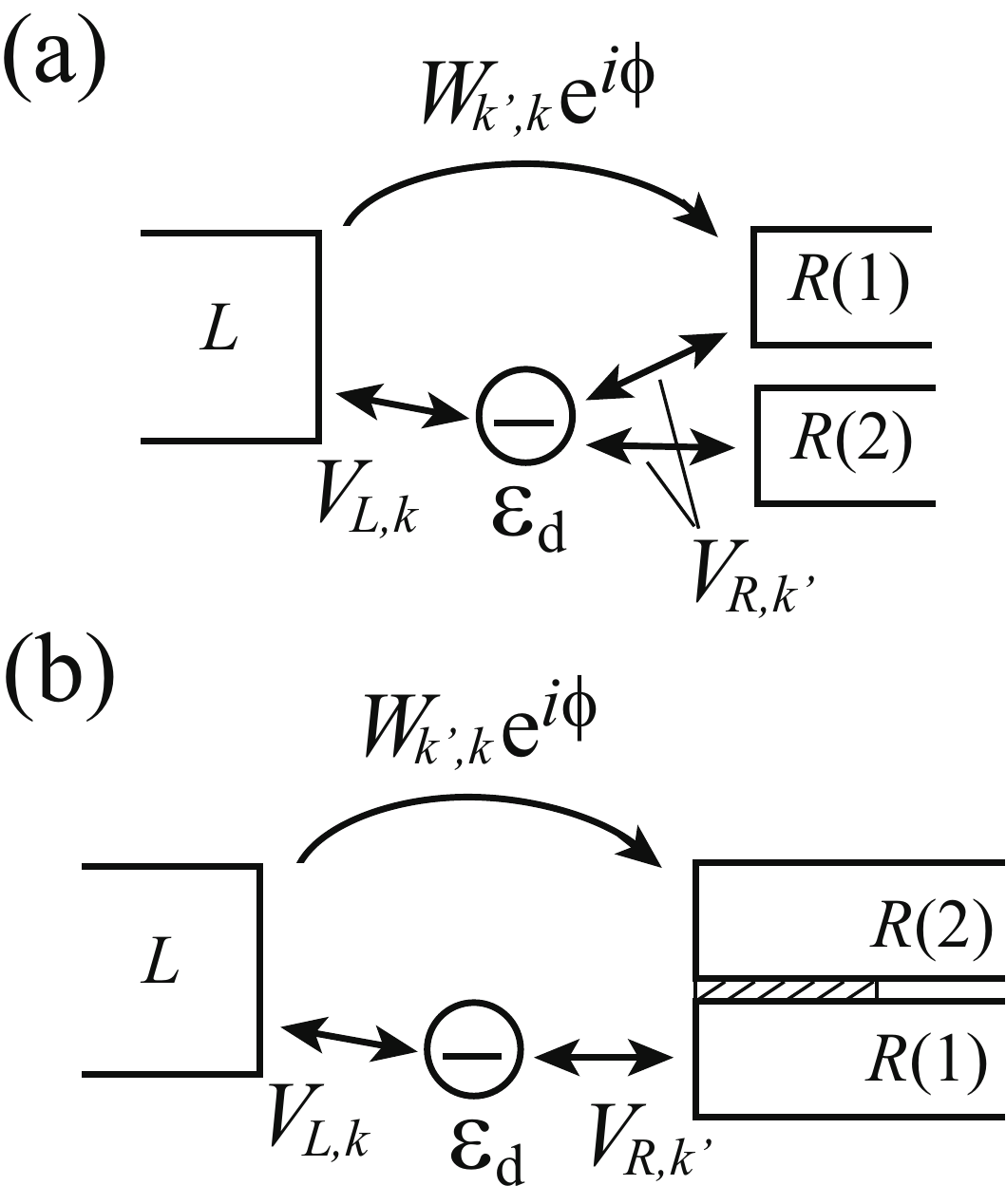}
\caption{%
Two specific models for the three-terminal system.
(a) Leads $L$ and $R(1)$ are connected to both the quantum dot and
upper arm of the ring, whereas lead $R(2)$ is connected to the
quantum dot only.
(b) Leads $R(1)$ and $R(2)$ are quantum wires which are tunnel-coupled
to each other at the hatched region.
Lead $L$ is connected to both the quantum dot and
upper arm of the ring, whereas lead $R(1)$ [$R(2)$] is connected to the
quantum dot [upper arm of the ring] only at the end of the leads.
}
\label{Fig4:model_3term}
\end{figure}

In subsections IV.B and C,
we examine two specific models depicted in Fig.\ \ref{Fig4:model_3term}.
In Fig.\ \ref{Fig4:model_3term}(a), leads $L$ and $R(1)$ are connected
to both the QD
and upper arm of  the ring, whereas lead $R(2)$ is connected to the QD
only. We vary the strength of tunnel coupling to lead $R(2)$,
$\Gamma_R^{(2)}$, to investigate a crossover from two- to three-terminal
phase measurement. In Fig.\ \ref{Fig4:model_3term}(b),
we model the experimental situation by Takada {\it et al}.,
in which leads $R(1)$ and $R(2)$ are partly-coupled quantum wires
\cite{PhysRevLett.113.126601,doi:10.1063/1.4928035,PhysRevB.94.081303}.

\subsection{Measured phase shift for $U=0$}

For the three-terminal model in Fig.\ \ref{Fig1:model0}(b) with leads
$L$, $R(1)$, and $R(2)$,
we introduce the following dimensionless parameters:
\begin{equation}
\gamma_{R}^{(j)}=\frac{\Gamma_{R}^{(j)}}{\Gamma_{R}},
\ \
y_{R}^{(j)} = \frac{x_{R}^{(j)}}{x_{R}},
\ \
q_{R}^{(j)}=
\frac{
\sqrt{\Gamma_{R}^{(j)} x_{R}^{(j)}} p_{R}^{(j)}}
{\sqrt{\Gamma_{R} x_{R}} p_{R}}
\label{eq:ratio_of_R(j)}
\end{equation}
for $j=1$ and $2$. They are the ratios of contribution from
lead $R(j)$ to $\Gamma_{R}$, $x_{R}$, and
$\sqrt{\Gamma_{R} x_{R}} p_{R}$, respectively, and satisfy
the relations of
$\gamma_{R}^{(1)}+\gamma_{R}^{(2)}=y_{R}^{(1)}+y_{R}^{(2)}
=q_{R}^{(1)}+q_{R}^{(2)}=1$.

In the absence of $U$, Eqs.\ (\ref{eq:3terminal_IRa}) and
(\ref{eq:3terminal_IRb}) yield the conductance in the form of
\begin{eqnarray}
G^{(1)} &=&
\frac{2e^2}{h}
\frac{1}{[E_{\rm F}-\tilde{\varepsilon}_d(\phi)]^2+\tilde{\Gamma}^2}
\nonumber \\
& \times &
\left[
\frac{8\sqrt{\Gamma_{L} \Gamma_{R}x} p_L p_R}{(1+x)^2} F(\phi)
+(\mbox{$\phi$-indep.\ terms})
\right],
\label{eq:G(1)_U0a}
\end{eqnarray}
where
\begin{equation}
F(\phi)=q_{R}^{(1)}(\varepsilon-\varepsilon_d) \cos \phi 
+\left[ x(q_{R}^{(1)}-y_{R}^{(1)}) \Gamma_L (1-p_L^2)
+(\gamma_{R}^{(1)}-q_{R}^{(1)}) \Gamma_R \right]
\sin \phi.
\label{eq:G(1)_U0b}
\end{equation}
If we neglect the $\phi$-dependence in $\tilde{\varepsilon}_d(\phi)$ in
the denominator in Eq.\ (\ref{eq:G(1)_U0a}), the measured phase
$\phi_{\rm max}$ is given by
\begin{equation}
\tan \phi_{\rm max}
=\frac{x(y_{R}^{(1)}-q_{R}^{(1)}) \Gamma_L(1-p_L^2)
+(q_{R}^{(1)}-\gamma_{R}^{(1)})\Gamma_R}{q_{R}^{(1)}(\Gamma_L+\Gamma_R)}
\tan \theta_{\rm QD}^{(0)},
\label{eq:measured_phase}
\end{equation}
where $\theta_{\rm QD}^{(0)}$ is defined in Eq.\ (\ref{eq:def_theta_QD(0)}).
This is an approximate formula for the relation between the measured value
and intrinsic value of the transmission phase shift through the QD.

In the two-terminal geometry, lead $R(2)$ is absent and thus
$\gamma_{R}^{(1)}=y_{R}^{(1)}=q_{R}^{(1)}=1$. Then Eq.\
(\ref{eq:measured_phase}) yields $\tan \phi_{\rm max}=0$, i.e.,
$\phi_{\rm max}=0$ or $\pi$
in accordance with the Onsager's reciprocal theorem.

\begin{figure}[t]%
\centering
\includegraphics*[width=0.75\linewidth]{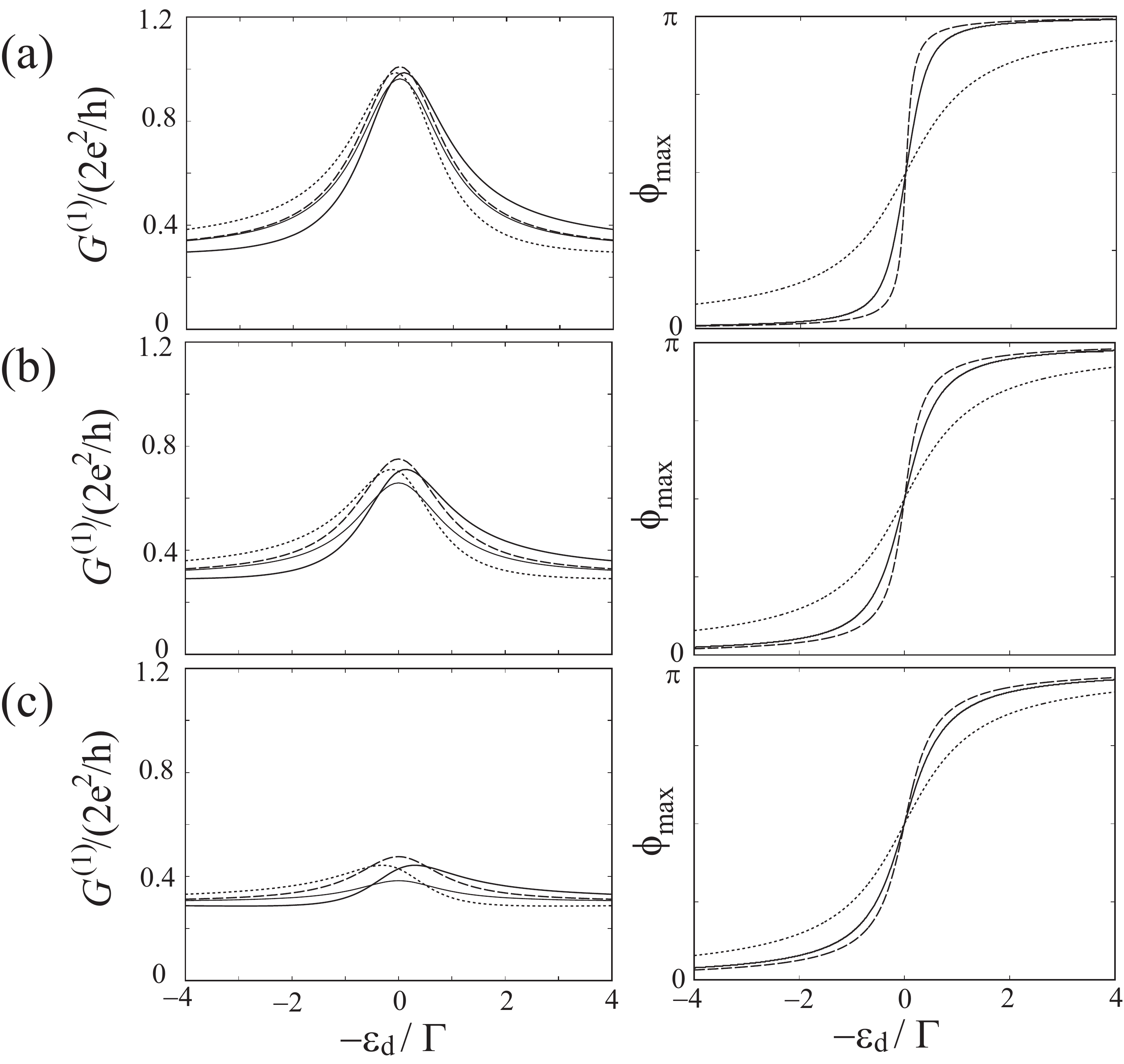}
\caption{%
Calculated results for the three-terminal model depicted in
Fig.\ \ref{Fig4:model_3term}(a) in the absence of $U$.
In the left panels, the conductance $G^{(1)}$ to lead $R(1)$
at temperature $T=0$ is plotted as a function of energy level
$\varepsilon_d$ in the quantum dot.
$\Gamma_L=\Gamma_R=\Gamma/2$, $x=0.09$ ($x_L=x_R=0.3$),
and $p_L=p_R=0.5$. The tunnel coupling to lead $R(2)$ is
increased from (a) to (c): 
(a) $\Gamma_R^{(2)}/\Gamma_R=0.2$, (b) 0.5, and (c) 0.8
with $\Gamma_R^{(1)}+\Gamma_R^{(2)}=\Gamma_R$.
The AB phase for the magnetic flux penetrating the ring is
$\phi=0$ (solid line), $\phi=\pi/2$ (broken line),
$\phi=\pi$ (dotted line), and $\phi=-\pi/2$ (thin solid line).
In the right panels, the measured phase shift $\phi_{\rm max}$
is plotted as a function of $\varepsilon_d$ (solid line),
which is numerically evaluated as the AB phase when $G^{(1)}(\phi)$
is maximal. $\phi_{\rm max}$ given by the formula in Eq.\
(\ref{eq:measured_phase_a}) is plotted by broken line,
whereas the transmission phase shift $\theta_{\rm QD}^{(0)}$ through
the quantum dot without the upper arm of the ring is plotted by dotted line.
}
\label{Fig5:3term_UOa}
\end{figure}

\subsection{Model in Fig.\ \ref{Fig4:model_3term}(a) with $U=0$}

To elucidate a crossover from two- to three-terminal measurement of the
transmission phase shift through the QD, we examine the model depicted in
Fig.\ \ref{Fig4:model_3term}(a) with $U=0$.
In this model, leads $L$ and $R(1)$ are connected to both the QD
and upper arm of  the ring, whereas lead $R(2)$ is connected to the QD
only. From $x_R^{(2)}=0$ and $\Gamma_R=\Gamma_R^{(1)}+\Gamma_R^{(2)}$,
dimensionless parameters in the previous subsection become
$\gamma_{R}^{(1)}=\Gamma_{R}^{(1)}/\Gamma_R$ and $y_{R}^{(1)} =q_{R}^{(1)}=1$.
In the QD, the effective energy level and linewidth are
$\tilde{\varepsilon}_d(\phi)=\varepsilon_d-
2 \sqrt{\Gamma_L \Gamma_R^{(1)}x} p_L p_R^{(1)} \cos\phi/(1+x)$ and
$\tilde{\Gamma} =\Gamma_L [1- x p_L^2/(1+x)] + 
\Gamma_R^{(1)} [1- x p_R^{(1) 2}/(1+x)] +\Gamma_R^{(2)}$, respectively.
Equation (\ref{eq:measured_phase}) yields an approximate relation of
\begin{equation}
\tan \phi_{\rm max}
=\frac{\Gamma_R^{(2)}}{\Gamma_L+\Gamma_R^{(1)}+\Gamma_R^{(2)}}
\tan \theta_{\rm QD}^{(0)},
\label{eq:measured_phase_a}
\end{equation}
which indicates that the measured phase shift $\phi_{\rm max}$
approaches the intrinsic phase shift 
$\theta_{\rm QD}^{(0)}$ with an increase in $\Gamma_R^{(2)}$.

Figure \ref{Fig5:3term_UOa} presents the calculated results for the model
in Fig.\ \ref{Fig4:model_3term}(a).
In the left panels, the conductance $G^{(1)}$ to lead $R(1)$ at $T=0$
is plotted as a function of energy level $\varepsilon_d$ in the QD.
$\Gamma_L=\Gamma_R=\Gamma/2$ and
(a) $\Gamma_R^{(2)}/\Gamma_R=0.2$, (b) 0.5, and (c) 0.8.
For small $\Gamma_R^{(2)}/\Gamma_R$ [panel (a)], $G^{(1)}$ is almost
the same at $\phi=\pm \pi/2$ corresponding to the Onsager's
reciprocal theorem in the two-terminal system.
With increasing $\Gamma_R^{(2)}/\Gamma_R$ [panels (b) and (c)],
the deviation from the theorem becomes more prominent. The peak height of
$G^{(1)}$ is reduced by stronger tunnel coupling to lead $R(2)$.

The right panels in Fig.\ \ref{Fig5:3term_UOa}
show $\phi_{\rm max}$ that is numerically evaluated from $G^{(1)}(\phi)$,
as a function of $\varepsilon_d$ (solid lines).
The intrinsic phase shift $\theta_{\rm QD}^{(0)}$ in Eq.\
(\ref{eq:def_theta_QD(0)}) is plotted by dotted lines.
Broken lines show $\phi_{\rm max}$ in Eq.\ (\ref{eq:measured_phase_a}),
indicating that the formula is a good approximation to estimate
$\phi_{\rm max}$ from $\theta_{\rm QD}^{(0)}$.
In panel (a) with $\Gamma_R^{(2)}/\Gamma_R=0.2$, $\phi_{\rm max}$ changes
almost abruptly from zero to $\pi$ around $\varepsilon_d=E_{\rm F}=0$,
which is close to the behavior in the two-terminal system.
For larger $\Gamma_R^{(2)}/\Gamma_R$, $\phi_{\rm max}$ changes
more gradually with $\varepsilon_d$ and closer to the intrinsic
phase shift $\theta_{\rm QD}^{(0)}$ although $\phi_{\rm max}$ does not
go to $\theta_{\rm QD}^{(0)}$ as $\Gamma_R^{(2)}/\Gamma_R \rightarrow 1$
under the condition of $\Gamma_L=\Gamma_R$.

\begin{figure}[t]%
\centering
\includegraphics*[width=0.4\linewidth]{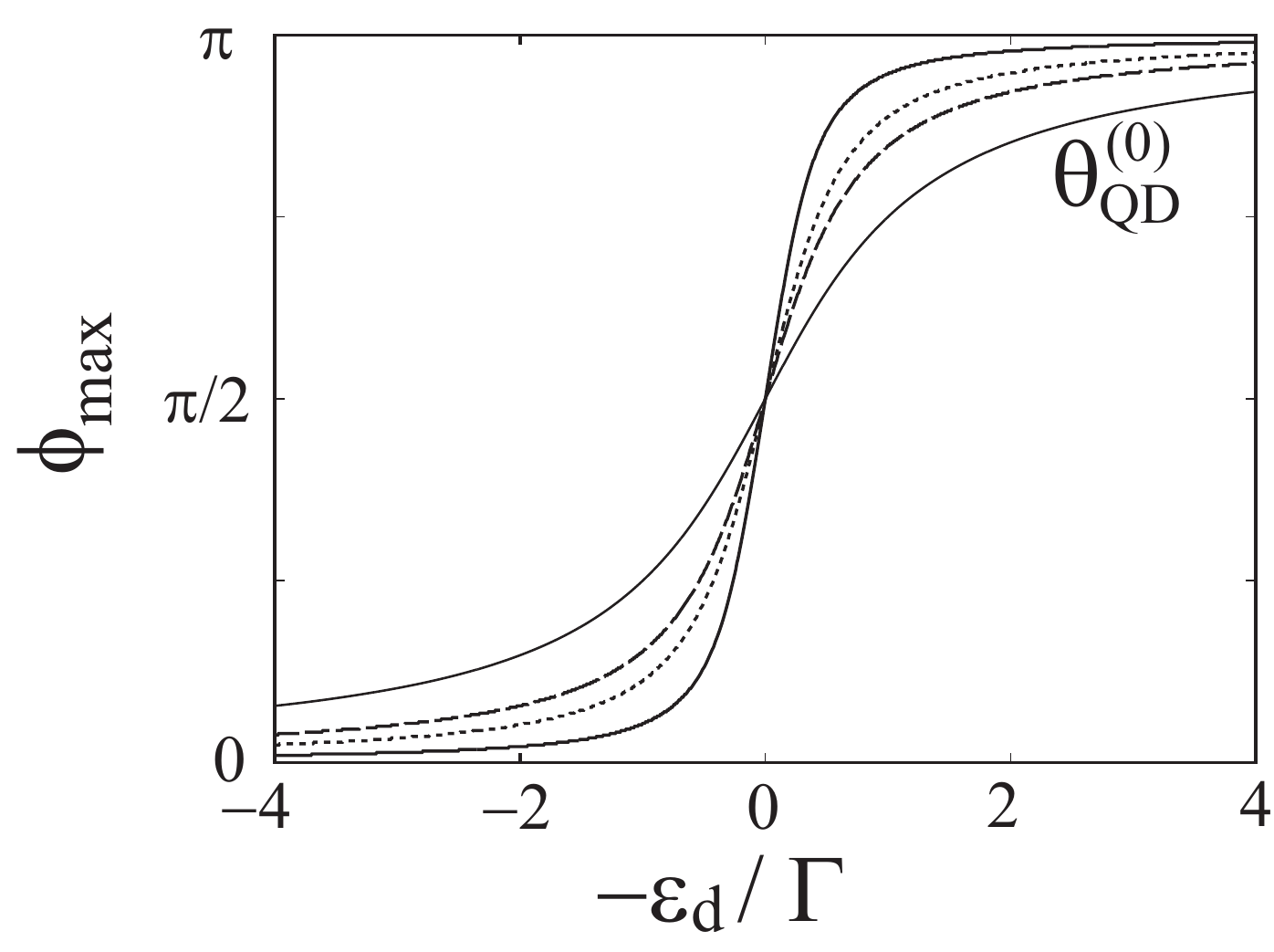}
\caption{%
Measured phase shift $\phi_{\rm max}$ as a function of energy level
$\varepsilon_d$ in the quantum dot, in the three-terminal model
depicted in Fig.\ \ref{Fig4:model_3term}(a) in the absence of $U$.
The data for $\phi_{\rm max}$ are the same as in the right panels
in Fig.\ \ref{Fig5:3term_UOa}.
$\Gamma_R^{(2)}/\Gamma_R=0.2$ (solid line), $0.5$ (dotted line),
and $0.8$ (broken line). A thin solid line indicates
the transmission phase shift $\theta_{\rm QD}^{(0)}$ through the
quantum dot without the upper arm of the ring.
}
\label{Fig6:3term_UOa2}
\end{figure}

To illustrate the crossover from the two- to three-terminal phase measurement,
we replot $\phi_{\rm max}$ for three values of
$\Gamma_R^{(2)}/\Gamma_R$ in a graph in Fig.\ \ref{Fig6:3term_UOa2}.

\subsection{Model in Fig.\ \ref{Fig4:model_3term}(b)}

Now we study the model shown in Fig.\ \ref{Fig4:model_3term}(b) to examine
the experimental situation using partly-coupled
quantum wires to form a mesoscopic ring
\cite{PhysRevLett.113.126601,doi:10.1063/1.4928035,PhysRevB.94.081303}.
We assume that leads $R(1)$ and $R(2)$ consist
of two equivalent wires $a$ and $b$ of single conduction channel.
They are tunnel-coupled to each other in the vicinity of their edges,
which mixes states $| a, k' \rangle$ in lead $a$ and
$| b, k' \rangle$ in lead $b$.
As a result, the edge states in leads $R(1)$ and $R(2)$ are given by
\begin{eqnarray}
| \psi_{R k'}^{(1)} \rangle =
 \alpha_R | a, k' \rangle + \beta_R | b, k' \rangle,
\label{eq:model_b1}
\\
| \psi_{R k'}^{(2)} \rangle =
\beta_R | a, k' \rangle - \alpha_R | b, k' \rangle,
\label{eq:model_b2}
\end{eqnarray}
respectively, with real coefficients $\alpha_R$ and $\beta_R$ 
($\alpha_R^2+\beta_R^2=1$). Far from the edges,
$| \psi_{R k'}^{(1)} \rangle \rightarrow | a, k' \rangle$ in
lead $R(1)$
and $| \psi_{R k'}^{(2)} \rangle \rightarrow | b, k' \rangle$
in lead $R(2)$ in an asymptotic way.

As shown in Fig.\ \ref{Fig4:model_3term}(b),
$| \psi_{R k'}^{(1)} \rangle$ in Eq.\
(\ref{eq:model_b1}) is coupled to the QD while
$| \psi_{R k'}^{(2)} \rangle$ in Eq.\ (\ref{eq:model_b2})
is connected to the upper arm of the ring. In the tunnel
Hamiltonian $H_T$ in Eq.\ (\ref{eq:tunnelH}), $V_{R,k'}=V_R \alpha_R$
and $\sqrt{w_{R,k'}}=\sqrt{w_R}\beta_R$ when state $k'$
belongs to lead $R(1)$ while $V_{R,k'}=V_R \beta_R$
and $\sqrt{w_{R,k'}}=-\sqrt{w_R}\alpha_R$
when state $k'$ belongs to lead $R(2)$.
Thus $\Gamma_R^{(1)}=\alpha_R^2 \Gamma_R$,
$\Gamma_R^{(2)}=\beta_R^2 \Gamma_R$, $x_R^{(1)}=\beta_R^2 x_R$,
and $x_R^{(2)}=\alpha_R^2 x_R$.

In this model, $p_R=0$ ($p_R^{(1)}=1$, $p_R^{(2)}=-1$) as
explained in Appendix C and in consequence
$\tilde{\varepsilon}_d(\phi)=\varepsilon_d$
in Eq.\ (\ref{eq:effective_e_d}). 
For $U=0$,
Eq.\ (\ref{eq:measured_phase}) exactly holds, which yields
\begin{equation}
\tan \phi_{\rm max}
=\frac{-x\Gamma_{L}(1-p_L^2)+\Gamma_{R}}{\Gamma_L+\Gamma_R}
\tan \theta_{\rm QD}^{(0)}.
\label{eq:measured_phase_b}
\end{equation}
Besides, the phase shift $\theta_{\rm QD}$ can be defined independently of
$\phi$, which satisfies the Friedel sum rule in the QD embedded in the ring
[see Eq.\ (\ref{eq:def_theta_QD}) in the case of $U=0$].
For both $U = 0$ and $U \ne 0$, we obtain an exact relation of
\begin{equation}
\tan \phi_{\rm max}
=\frac{-x\Gamma_{L}(1-p_L^2)+\Gamma_{R}}{\tilde{\Gamma}}
\tan \theta_{\rm QD},
\label{eq:measured_phase_b_Kondo}
\end{equation}
with $\tilde{\Gamma}=\Gamma_L [1- x p_L^2/(1+x)] + \Gamma_R$.
$\tan \theta_{\rm QD}=\tilde{\Gamma}/\varepsilon_d$ in the absence of
$U$ and $\tan \theta_{\rm QD}=\tilde{\Gamma}^*/\tilde{\varepsilon}_d^*$ in
the presence of $U$ when $E_{\rm F}=0$.
Neither Eq.\ (\ref{eq:measured_phase_b})
nor Eq.\ (\ref{eq:measured_phase_b_Kondo}) depend on $\alpha_R$ and
$\beta_R$.

\begin{figure}[t]%
\centering
\includegraphics*[width=0.45\linewidth]{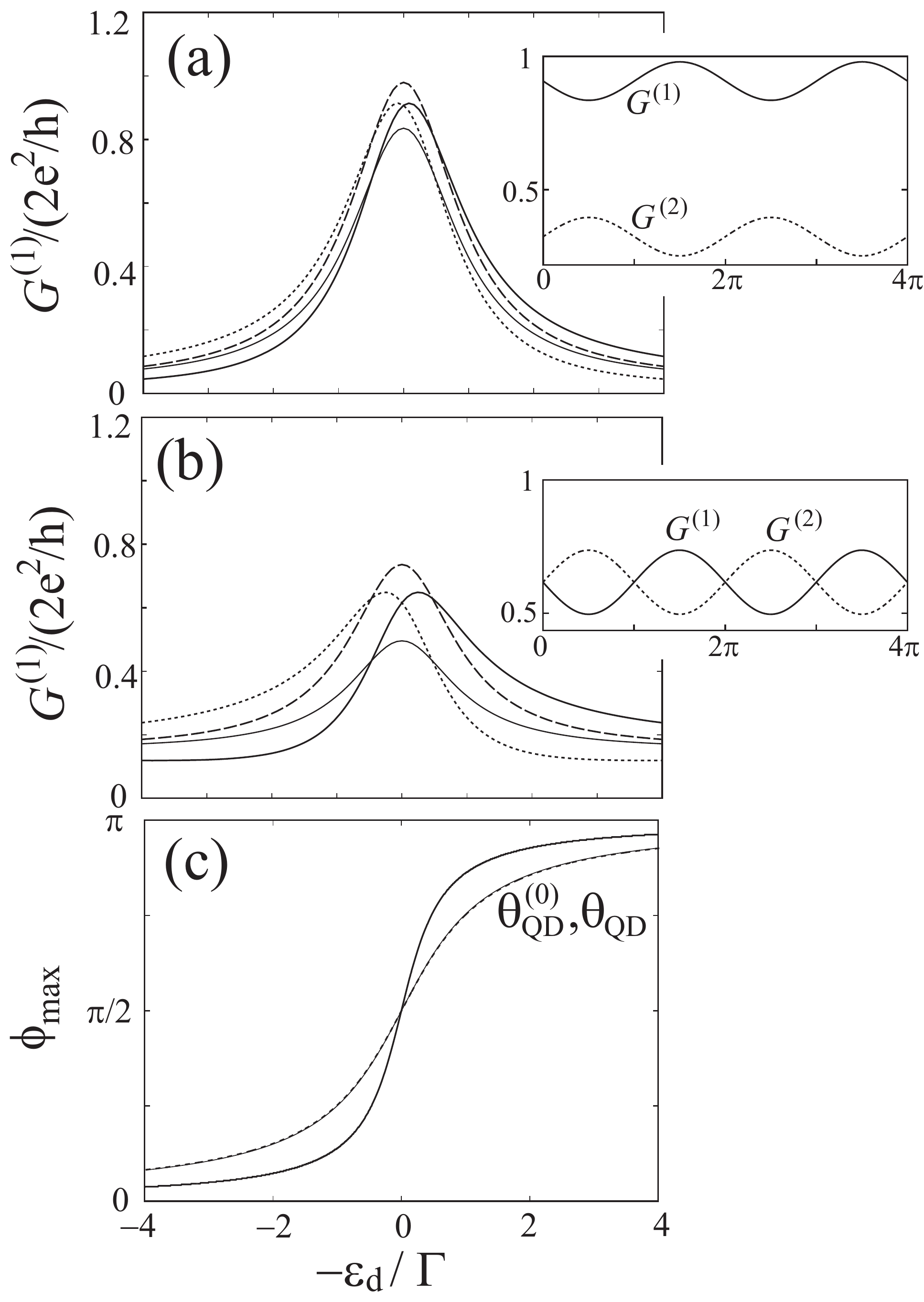}
\caption{%
Calculated results for the conductance and measured phase shift
in the three-terminal model depicted in Fig.\ \ref{Fig4:model_3term}(b)
in the absence of $U$. In the upper two panels,
the conductance $G^{(1)}$ to lead $R(1)$
at temperature $T=0$ is plotted as a function of energy level
$\varepsilon_d$ in the quantum dot.
$\Gamma_L=\Gamma_R=\Gamma/2$, $x=0.09$ ($x_L=x_R=0.3$),
and $p_L= 0.5$. (a) $\beta_R^2=0.1$, (b) $0.5$
($\alpha_R^2+\beta_R^2=1$; $\Gamma_R^{(1)}=\alpha_R^2 \Gamma_R$,
$\Gamma_R^{(2)}=\beta_R^2 \Gamma_R$, $x_R^{(1)}=\beta_R^2 x_R$,
and $x_R^{(2)}=\alpha_R^2 x_R$, see text).
The AB phase for the magnetic flux penetrating the ring is
$\phi=0$ (solid line), $\phi=\pi/2$ (broken line),
$\phi=\pi$ (dotted line), and $\phi=-\pi/2$ (thin solid line).
In panel (c), the measured phase shift $\phi_{\rm max}$
is plotted as a function of $\varepsilon_d$ (solid line),
which is defined by the AB phase when $G^{(1)}(\phi)$ is maximal.
$\phi_{\rm max}$ does not depend on $\beta_R$.
The phase shift $\theta_{\rm QD}^{(0)}$ through the QD without
the upper arm of the ring is plotted by dotted line, which is almost
overlapped by $\theta_{\rm QD}$ (thin solid line) that
satisfies the Friedel sum rule in the QD embedded in the ring.
Insets in panels (a) and (b): $G^{(1)}$ and $G^{(2)}$
[conductance to lead $R(2)$] as a function of the AB phase $\phi$,
at $\varepsilon_d=0$.
}
\label{Fig7:3term_UOb}
\end{figure}

We show the calculated results for $U=0$ in Fig.\ \ref{Fig7:3term_UOb}.
In panels (a) and (b), the conductance $G^{(1)}$ is shown
as a function of energy level $\varepsilon_d$ in the QD,
for (a) $\beta_R^2=0.1$ and (b) $0.5$.
The height of $G^{(1)}$ depends on $\phi$ more largely in
panel (b) than in panel (a) though $\phi_{\rm max}$ does not depend
on $\beta_R$.

Figure \ref{Fig7:3term_UOb}(c) plots $\phi_{\rm max}$ that is
numerically evaluated from $G^{(1)}(\phi)$.
It changes smoothly from zero to $\pi$
via $\pi/2$ at $\varepsilon_d=0$. $\phi_{\rm max}$
quantitatively deviates from $\theta_{\rm QD}^{(0)}$
and $\theta_{\rm QD}$ (dotted and thin solid lines).
Their relations are exactly given by Eqs.\ (\ref{eq:measured_phase_b}) and
(\ref{eq:measured_phase_b_Kondo}).

It should be mentioned that the sum of the currents to
leads $R(1)$ and $R(2)$, $I_R^{(1)}+I_R^{(2)}$, does not
depend on the AB phase $\phi$, reflecting $p_R=0$ in this model
(see Appendix C).
Therefore, the AB oscillation of $G^{(1)}(\phi)$ is
out-of-phase to that of $G^{(2)}(\phi)$, as indicated in the insets
in Fig.\ \ref{Fig7:3term_UOb}. $\phi_{\rm max}$
evaluated from $G^{(1)}$ behaves similarly to $\theta_{\rm QD}^{(0)}$,
while that from $G^{(2)}$ similarly to $-\theta_{\rm QD}^{(0)}$,
irrespective of the absence or presence of $U$.
This agrees with the experimental observation by Takada {\it et al}.\
\cite{PhysRevLett.113.126601,doi:10.1063/1.4928035}.

\begin{figure}[t]%
\centering
\includegraphics*[width=0.4\linewidth]{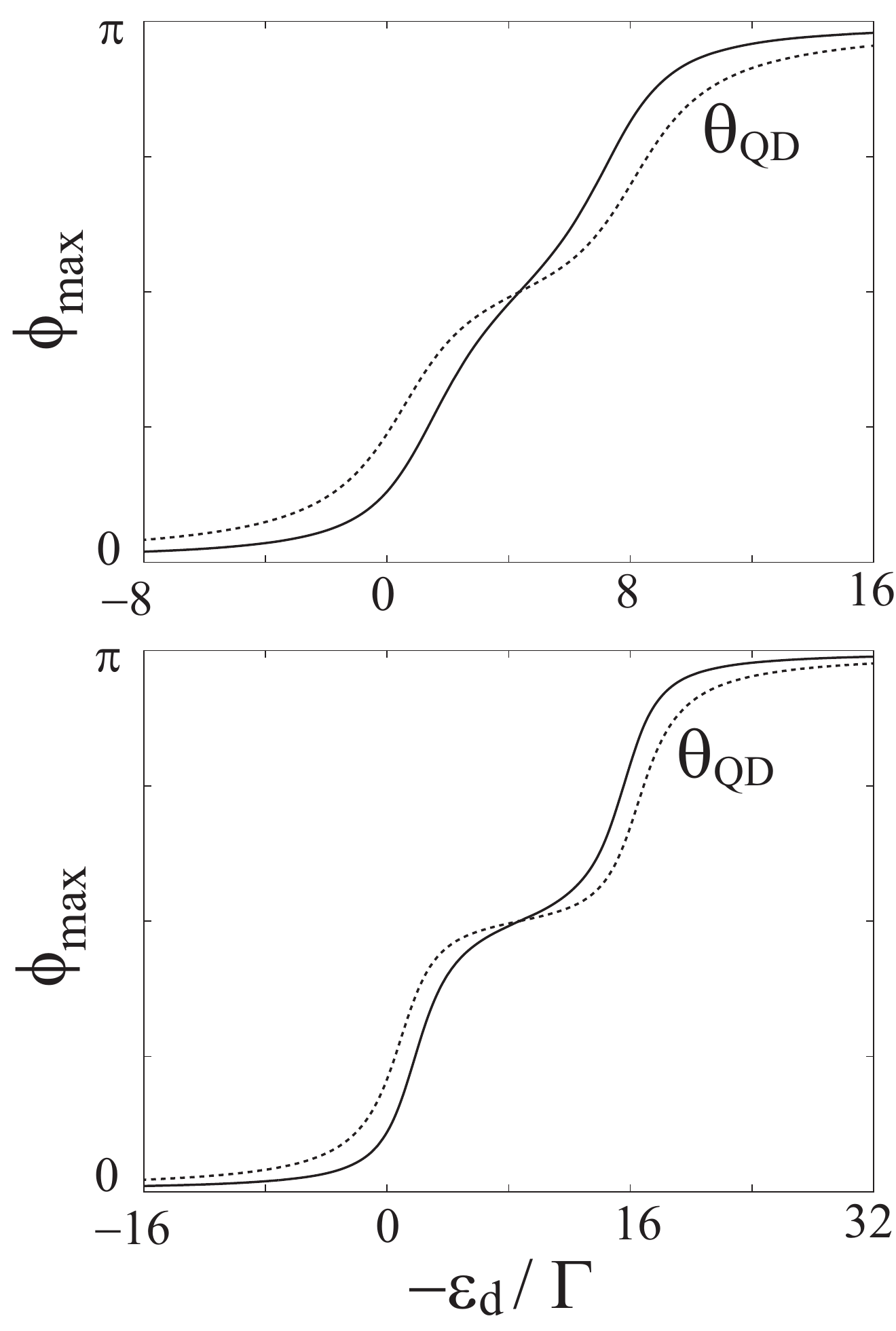}
\caption{%
Calculated results for the measured phase shift in the three-terminal model
depicted in Fig.\ \ref{Fig4:model_3term}(b) in the presence of $U$.
The measured phase shift $\phi_{\rm max}$ is plotted by solid line
as a function of energy level $\varepsilon_d$ in the quantum dot.
$\phi_{\rm max}$ is numerically evaluated as the AB phase
at which the conductance $G^{(1)}(\phi)$ to lead $R(1)$ is maximal
at temperature $T=0$.
$\Gamma_L=\Gamma_R=\Gamma/2$, $x=0.09$ ($x_L=x_R=0.3$),
and $p_L= 0.5$. (a) $U/\Gamma=8$ and (b) $16$.
$\theta_{\rm QD}$ calculated from the Friedel sum rule,
$\theta_{\rm QD}=\pi \langle n_{\sigma} \rangle$, is plotted
by dotted line. $\phi_{\rm max}$ and $\theta_{\rm QD}$ are
related to each other by Eq.\ (\ref{eq:measured_phase_b_Kondo}).
}
\label{Fig8:3term_Ub}
\end{figure}

Finally, the measured phase is discussed in the Kondo regime with
$U \ne 0$. In Fig.\ \ref{Fig8:3term_Ub}, we plot $\phi_{\rm max}$ that
is numerically evaluated from $G^{(1)}$, as a function of energy level
$\varepsilon_d$ in the QD.
(a) $U/\Gamma=8$ and (b) $16$ with $\Gamma_L=\Gamma_R=\Gamma/2$.
In the Kondo valley ($-U < \varepsilon_d < 0$), the phase locking at
$\pi/2$ is observable by a ``double-slit experiment'' using
the QD interferometer.
We calculate the intrinsic phase shift $\theta_{\rm QD}$ using the Friedel
sum rule $\theta_{\rm QD}=\pi \langle n_{\sigma} \rangle$, where
$\langle n_{\sigma} \rangle$ is given by the Bethe ansatz exact solution
(dotted line). $\phi_{\rm max}$ and $\theta_{\rm QD}$ are related
to each other by Eq.\ (\ref{eq:measured_phase_b_Kondo}).
The phase locking seems smeared in the curve of the measured phase shift
$\phi_{\rm max}$,
in comparison with the intrinsic phase shift $\theta_{\rm QD}$.

\section{Discussion}

In our models shown in Figs.\ \ref{Fig1:model0}(a) and (b), we
assume a separable form for the tunnel coupling between the leads in
Eq.\ (\ref{eq:Wseparable}). Here, we discuss the justification of this form
using a tight-binding model.
We also show that $|p_{\alpha}|<1$ in the presence of multiple conduction
channels in lead $\alpha$.

\begin{figure}[t]%
\centering
\includegraphics*[width=0.5\linewidth]{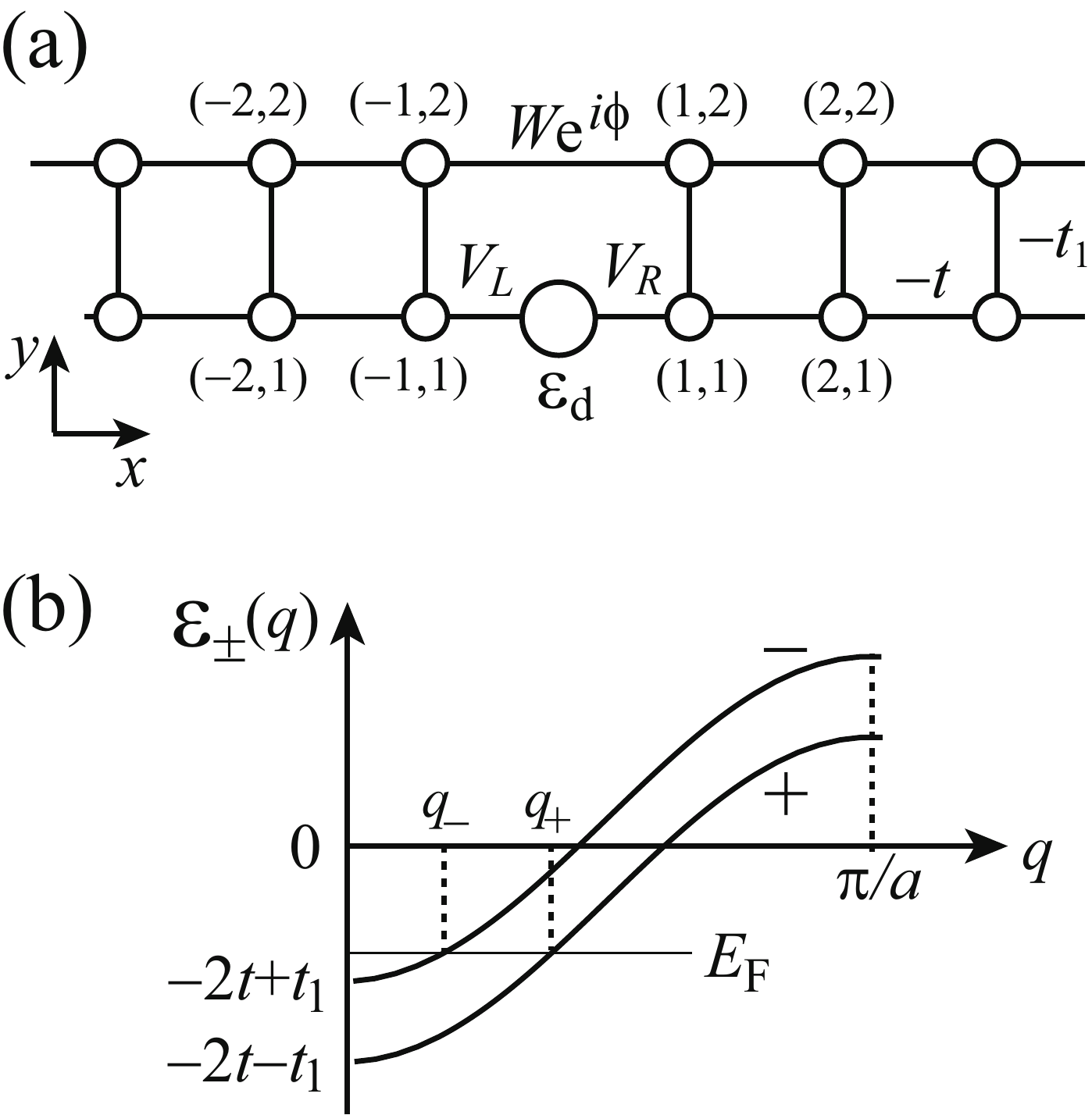}
\caption{%
(a) A tight-binding model for the quantum-dot (QD) interferometer.
A QD is connected to sites $(-1,1)$ and $(1,1)$ by transfer
integrals $V_L$ and $V_R$, respectively, whereas the upper arm of the ring
couples sites $(-1,2)$ and $(1,2)$ by $We^{\pm i\phi}$ in $\pm x$ direction,
where $\phi$ is the AB phase for the magnetic flux penetrating the ring
(central rectangular region including the QD). Leads $L$ and $R$ consist of
two sites in width ($y$ direction) and $N$ sites in length
($x$ direction; $N \gg 1$), in which the transfer integral is
$-t$ ($-t_1$) in the $x$ ($y$) direction and
the lattice constant is $a$. (b) Two subbands in the leads, 
$\varepsilon_{\pm}(q)=\mp t_1 -2t \cos qa$, as a function of wavenumber
$q$ in the $x$ direction ($0<q<\pi/a$).
There are two conduction channels when the Fermi
level $E_{\rm F}$ intersects both the subbands at $q=q_{\pm}$.
}
\label{Fig9:tight_binding}
\end{figure}

As a simple example, let us consider the model depicted in Fig.\
\ref{Fig9:tight_binding}(a).
The leads consist of two sites in width and $N$ sites in length ($N \gg 1$).
The eigenvalues of the Hamiltonian for leads $L$ and $R$ form two
subbands $\varepsilon_{\pm}(q)$, where $q$ is the wavenumber in the $x$
direction ($0<q<\pi/a$) with $a$ being the lattice constant
[Fig.\ \ref{Fig9:tight_binding}(b)]. The corresponding states are
\begin{eqnarray}
|L;q,\pm \rangle
&=&
\frac{-1}{\sqrt{N+1}}
\sum_{j=-N}^{-1} \left( |j,1 \rangle \pm |j,2 \rangle \right)
\sin qja,
\label{eq:TightBindingModel1}
\\
|R;q,\pm \rangle
&=&
\frac{1}{\sqrt{N+1}}
\sum_{j=1}^{N} \left( |j,1 \rangle \pm |j,2 \rangle \right)
\sin qja,
\label{eq:TightBindingModel2}
\end{eqnarray}
where $|j,\ell \rangle$ is the Wannier function at site $(j,\ell)$.
The tunnel coupling between $|L;q,\gamma \rangle$ and
$|R;q',\gamma' \rangle$ ($\gamma$, $\gamma'=\pm$) is expressed as
$W_{q',\gamma'; q,\gamma}=\psi_{R; q',\gamma'}(1,2)
W \psi_{L; q,\gamma}(-1,2)$
using the wavefunctions at the edge of the leads,
$\psi_{L; q,\pm}(-1,2)=\langle -1,2 | L;q,\pm \rangle$ and 
$\psi_{R; q',\pm}(1,2)=\langle 1,2 | R;q',\pm \rangle$.
In consequence $W_{q',\gamma'; q,\gamma}$
has a separable form in Eq.\ (\ref{eq:Wseparable}) with
\begin{eqnarray}
\sqrt{w_{L; q,\gamma}} &=& \sqrt{W}\psi_{L; q,\gamma}(-1,2),
\label{eq:TightBindingModel3}
\\
\sqrt{w_{R; q',\gamma'}} &=& \sqrt{W}\psi_{R; q',\gamma'}(1,2).
\label{eq:TightBindingModel4}
\end{eqnarray}

When the Fermi level intersects both the subbands,
there are two conduction channels, labeled by $k=(q,\pm)$,
as indicated in Fig.\ \ref{Fig9:tight_binding}(b). Then
\begin{equation}
p_L=p_R=\frac{\sin q_+a - \sin q_-a}{\sin q_+a + \sin q_-a},
\label{eq:TightBindingModel5}
\end{equation}
where $q_{\pm}$ are the intersections between the subband
$\pm$ and Fermi level, as derived in Appendix D.
Thus $|p_{L,R}|<1$. 
On the other hand, $p_{L,R}=\pm 1$, in the case of single conduction
channel when $E_{\rm F}$ crosses one of the subbands.

Although we have considered a specific model in
Fig.\ \ref{Fig9:tight_binding}(a),
the separable form of $W_{k',k}$ in Eq.\ (\ref{eq:Wseparable})
should be justified when the system is described by a
tight-binding model in general. Then $\sqrt{w_{L,k}}$ ($\sqrt{w_{R,k'}}$)
is proportional to the wavefunction
$\psi_{L,k}$ ($\psi_{R,k'}$) at the edge of the lead,
as in Eqs.\ (\ref{eq:TightBindingModel3}) and (\ref{eq:TightBindingModel4}).
We could also claim that $p_{L,R}<1$
for the leads of multiple conduction channels and $p_{L,R}=1$ for the
leads of single channel in usual cases. Precisely speaking,
the presence of multiple channels is a necessary condition
for $p_{L,R}<1$: $p_{\alpha}$ is determined by the detailed shape of
the system around a junction between the ring and lead $\alpha$
through Eq.\ (\ref{eq:p_overlap_integral}).

We comment on the generality of our models. In this section, we have examined a
model in which the subbands ($\pm$) are well defined in the leads.
Then the state in the leads is labeled by $k=(q,\pm)$ in the presence of
two conduction channels.
This is not the case in experimental systems of various shape. We believe
that $\Gamma_{\alpha}$, $x_{\alpha}$, and $p_{\alpha}$ can be defined in
Eqs.\ (\ref{eq:Gamma})--(\ref{eq:introduce_p}) using
state-dependent tunnel couplings without loss of generality.
In our models in Figs.\ \ref{Fig1:model0}(a) and (b),
we assume a single conduction channel in the upper arm of the ring.
The multiple channels in the arm should be beyond the scope of our study.

\section{Conclusions}

We have theoretically examined the transport through an Aharonov-Bohm ring
with an embedded quantum dot (QD), the so-called QD interferometer,
to address two controversial issues,
one concerns the shape of the conductance peak as a function of energy
level $\varepsilon_d$ in the QD and the other is about
the phase measurement in the multi-terminal geometry
as a double-slit experiment. For the purpose, we have generalized
a previous model in Refs.\ \cite{PhysRevLett.86.5128,PhysRevLett.87.156803}
to consider multiple conduction channels in leads $L$ and $R$.
In our model, the tunnel couplings between the QD and leads and that
between the leads depend on the states in the leads, as shown in
Figs.\ \ref{Fig1:model0}(a) and (b). This gives rise to a parameter
$p_{\alpha}$ ($|p_{\alpha}| \le 1$) to characterize a connection
between the two arms of the ring through lead $\alpha$ ($=L$, $R$),
which is equal to the overlap integral between the conduction modes
coupled to the upper and lower arms of the ring.

First, we have examined the shape of the conductance peak
in the two-terminal geometry,
in the absence of electron-electron interaction $U$ in the QD.
We have shown an asymmetric Fano resonance at $|p_{L,R}| \approx 1$
and an almost symmetric Breit-Wigner resonance at $|p_{L,R}| < 0.5$.
Hence our model could explain the experimental results of both an
asymmetric Fano resonance \cite{PhysRevLett.88.256806} and almost symmetric
Breit-Wigner resonance \cite{PhysRevLett.113.126601}, with fitting parameters
$p_{L,R}$ to their data.

Second, we have taken into account the Kondo effect in the presence of $U$,
using the Bethe ansatz exact solution, and precisely evaluated the conductance
at temperature $T=0$. We have shown a crossover from an asymmetric Fano-Kondo
resonance \cite{PhysRevLett.87.156803} to the Kondo plateau with
changing $p_{L,R}$.

Our model is also applicable to the multi-terminal geometry to address the
second issue on the measurement of the transmission phase shift through
the QD by a double-slit experiment.
We have studied the measured phase $\phi_{\rm max}$, the AB phase at
which the conductance $G^{(1)}(\phi)$ to lead $R(1)$ is maximal
in Fig.\ \ref{Fig1:model0}(b). In the absence of $U$,
Eq.\ (\ref{eq:measured_phase}) indicates the relation of
$\phi_{\rm max}$ to an intrinsic phase shift $\theta_{\rm QD}^{(0)}$
that is the phase shift through the QD without the upper arm of
the ring.
We have examined two specific models in the three-terminal geometry,
depicted in Fig.\ \ref{Fig4:model_3term}.
We have discussed a crossover from two- to three-terminal phase
measurement in the former and simulated the experimental system consisting of
two quantum wires
\cite{PhysRevLett.113.126601,doi:10.1063/1.4928035,PhysRevB.94.081303}
in the latter. Using the latter model,
we have shown how precisely the phase locking at $\pi/2$ is measured in the
Kondo regime.

\begin{acknowledgments}
We appreciate fruitful discussions with Dr.\ Akira Oguri.
This work was partially supported by JSPS KAKENHI Grant Numbers
JP26220711, JP15H05870, and JST-CREST Grant Number JPMJCR1876.
\end{acknowledgments}

\appendix

\section{Current formulation using Keldysh Green's functions}

The current is formulated for the multi-terminal model depicted
in Fig.\ \ref{Fig1:model0}(b), using the Keldysh Green's functions
\cite{PhysRevLett.68.2512,PhysRevB.50.5528,Jauho2008}.
The chemical potential in lead $L(j)$ [$R(j)$] is denoted by
$\mu_L^{(j)}$ [$\mu_R^{(j)}$].
The spin index $\sigma$ is omitted in this appendix.

\subsection{Keldysh Green's functions}
The retarded, advanced, and lesser Green's functions are defined by
\begin{eqnarray}
G_{d,Lk}^{\rm r}(t,t') &=&
\frac{1}{i\hbar} \langle \{ d(t), a_{Lk}^{\dagger}(t')  \} \rangle
\theta(t-t'),
\\
G_{d,Lk}^{\rm a}(t,t') &=&
-\frac{1}{i\hbar} \langle \{ d(t), a_{Lk}^{\dagger}(t')  \} \rangle
\theta(t'-t),
\\
G_{d,Lk}^{<}(t,t') &=&
\frac{-1}{i\hbar} \langle a_{Lk}^{\dagger}(t') d(t) \rangle,
\end{eqnarray}
respectively, where $\{ A, B\}=AB+BA$ and $\theta(t)$ is the
Heaviside step function.
The other Green's functions,
$G_{d,d}^{\lambda}$, $G_{Lk,Rk'}^{\lambda}$, etc.\
($\lambda=$ r, a, $<$), are defined in a similar manner.
The average is taken for the stationary state and
hence all the Green's functions depend on $t-t'$ only.
Note that $G_{d,Lk}^{<}(t-t')=-[G_{Lk,d}^{<}(t'-t)]^*$
and $G_{d,d}^{<}(t-t')=-[G_{d,d}^{<}(t'-t)]^*$. The Fourier
transformation ($t-t' \rightarrow \omega$) yields
$G_{d,Lk}^{<}(\omega)=-[G_{Lk,d}^{<}(\omega)]^*$ and
$G_{d,d}^{<}(\omega)=-[G_{d,d}^{<}(\omega)]^*$.

We also introduce the Green's functions in isolate leads $L$ and $R$,
in the absence of tunnel coupling, $H_T$ in Eq.\ (\ref{eq:tunnelH}).
For example,
\begin{eqnarray}
g_{Lk}^{\rm r}(t,t') &=&
\frac{1}{i\hbar} \langle \{ a_{Lk}(t), a_{Lk}^{\dagger}(t')  \} \rangle
\theta(t-t')=\frac{1}{i\hbar} e^{-i \varepsilon_k (t-t')/\hbar}\theta(t-t'),
\\
g_{Lk}^{<}(t,t') &=&
\frac{-1}{i\hbar} \langle a_{Lk}^{\dagger}(t') a_{Lk}(t) \rangle
=\frac{-1}{i\hbar} f_{L}^{(j)}(\varepsilon_k)
 e^{-i \varepsilon_k (t-t')/\hbar},
\end{eqnarray}
where $f_{L}^{(j)}(\varepsilon)=
[(\varepsilon-\mu_L^{(j)})/(k_{\rm B}T)+1]^{-1}$ is the
Fermi distribution function in lead $L(j)$ that state $k$
belongs to ($j=1$ or $2$). The Fourier transformation leads to
\begin{eqnarray}
g_{Lk}^{\rm r}(\omega) &=& \frac{1}{\hbar\omega-\varepsilon_k+i\delta}
\nonumber \\
&=&
P\frac{1}{\hbar\omega-\varepsilon_k}-i\pi \delta(\hbar\omega-\varepsilon_k),
\\
g_{Lk}^{<}(\omega) &=& 2\pi i f_{L}^{(j)}(\hbar\omega)
\delta(\hbar\omega-\varepsilon_k).
\end{eqnarray}
In the following calculations, the real part (principal value) of 
$g_{\alpha k}^{\rm r}(\omega)$
and $g_{\alpha k}^{\rm a}(\omega)=[g_{\alpha k}^{\rm r}(\omega)]^*$
is disregarded in the summation over $k$, assuming a wide band limit.

In the next subsection, $G_{d,Lk}^{<}$ is replaced by $G_{d,d}^{\rm r}$
and $G_{d,d}^{<}$. For this purpose, their relation is derived
in the following. In the Baym-Kadanoff-Keldysh nonequilibrium
techniques, a complex-time contour is considered from $t=-\infty$ to
$t=t_0$ just above the real axis and from $t=t_0$ to $t=-\infty$ just
below the real axis. For the contour-ordered Green's function,
\begin{equation}
G_{d,Lk}^{\rm C}(\tau,\tau')=\frac{1}{i\hbar}
\langle \mathcal{T}_{\rm C} d(\tau) a_{Lk}^{\dagger}(\tau')  \rangle,
\end{equation}
the equation-of-motion method yields \cite{PhysRevB.50.5528,Jauho2008}
\begin{equation}
G_{d,Lk}^{\rm C}(\tau,\tau') = \int d \tau_1
\left[
G_{d,d}^{\rm C}(\tau,\tau_1) V_{Lk} +
\sum_{k'}^{(1),(2)} G_{d,Rk'}^{\rm C}(\tau,\tau_1)
W_{k',k}e^{i\phi} \right] g_{Lk}^{\rm C}(\tau_1,\tau').
\end{equation}
According to the Langreth's theorem \cite{Jauho2008,grandstrand:2004},
this results in
\begin{equation}
G_{d,Lk}^{\rm r}(t,t') = \int d t_1
\left[
G_{d,d}^{\rm r}(t,t_1) V_{Lk} +
\sum_{k'}^{(1),(2)} G_{d,Rk'}^{\rm r}(t,t_1)
W_{k',k}e^{i\phi} \right] g_{Lk}^{\rm r}(t_1,t'),
\label{eq:Langreth1}
\end{equation}
and
\begin{eqnarray}
G_{d,Lk}^<(t,t') &=& \int d t_1
\Biggl\{
\left[
G_{d,d}^{\rm r}(t,t_1) V_{Lk} +
\sum_{k'}^{(1),(2)} G_{d,Rk'}^{\rm r}(t,t_1)
W_{k',k}e^{i\phi} \right] g_{Lk}^<(t_1,t')
\nonumber \\
& &
+ \left[
G_{d,d}^<(t,t_1) V_{Lk} +
\sum_{k'}^{(1),(2)} G_{d,Rk'}^<(t,t_1)
W_{k',k}e^{i\phi} \right] g_{Lk}^{\rm a}(t_1,t')
\Biggr\}.
\label{eq:Langreth2}
\end{eqnarray}
Similar relations are obtained for $G_{d,Rk'}^{\rm r}$, etc.

\subsection{Current formula using $G_{d,d}^{\rm r}$ and $G_{d,d}^<$}

We express the current from lead $L(1)$ in terms of
$G_{d,d}^{\rm r}$ and $G_{d,d}^<$.
The substitution of the Hamiltonian in Eq.\ (\ref{eq:Hamiltonian})
into Eq.\ (\ref{eq:current_L^(1)}) results in
\begin{equation}
I_L^{(1)} = 
-\frac{2e}{i\hbar} \sum_{k}^{(1)}
\left[
V_{Lk}\langle a_{Lk}^{\dagger} d - d^{\dagger} a_{Lk} \rangle
+ \sum_{k'}^{(1),(2)} W_{k',k}
\langle e^{-i\phi}a_{Lk}^{\dagger} a_{Rk'} - 
        e^{i\phi}a_{Rk'}^{\dagger} a_{Lk} \rangle
\right].
\end{equation}
We have added a factor of two by the summation over spin index $\sigma$.
This equation is rewritten as
\begin{eqnarray}
I_L^{(1)} &=& 4e {\rm Re}
\sum_{k}^{(1)}
\left[
V_{Lk} G_{d,Lk}^{<}(t,t)
+\sum_{k'}^{(1),(2)} W_{k',k}e^{-i\phi}G_{Rk',Lk}^{<}(t,t)
\right]
\nonumber \\
&=&
\frac{4e}{2\pi} {\rm Re} \int d\omega
\sum_{k}^{(1)}
\left[
V_{Lk} G_{d,Lk}^{<}(\omega)
+\sum_{k'}^{(1),(2)} W_{k',k}e^{-i\phi}G_{Rk',Lk}^{<}(\omega)
\right].
\end{eqnarray}
Hence we need to calculate two terms in the integral,
\begin{eqnarray}
X_0 &=& \sum_{k}^{(1)} V_{Lk} G_{d,Lk}^{<}(\omega),
\\
Y_0 &=& \sum_{k}^{(1)}
      \sum_{k'}^{(1),(2)} W_{k',k}e^{-i\phi}G_{Rk',Lk}^{<}(\omega).
\end{eqnarray}

Let us consider $X_0$.
Using the Fourier transformation of Eq.\ (\ref{eq:Langreth2}),
we obtain
\begin{eqnarray}
X_0 &=& i \Gamma_L^{(1)} [2f_L^{(1)}(\hbar\omega)G_{d,d}^{\rm r}(\omega)
+G_{d,d}^<(\omega)]
\nonumber \\
& &
+i \tilde{p}_L^{(1)} e^{i\phi}  \sum_{k'}^{(1),(2)}\sqrt{w_{Rk'}}
[2f_L^{(1)}(\hbar\omega)G_{d,Rk'}^{\rm r}(\omega)
+G_{d,Rk'}^<(\omega)],
\end{eqnarray}
where $\tilde{p}_{\alpha}^{(j)}=\sqrt{\Gamma_{\alpha}^{(j)}
x_{\alpha}^{(j)}}p_{\alpha}^{(j)}$. Then we need
\begin{eqnarray}
X_{1} &=& \sum_{k'}^{(1),(2)}\sqrt{w_{Rk'}}G_{d,Rk'}^{\rm r}(\omega),
\\
X_{2} &=& \sum_{k'}^{(1),(2)}\sqrt{w_{Rk'}}G_{d,Rk'}^{<}(\omega).
\end{eqnarray}
For $X_1$, we use an equation for $G_{d,Rk'}^{\rm r}$ corresponding
to Eq.\ (\ref{eq:Langreth1}) for $G_{d,Lk}^{\rm r}$, which leads to
\begin{equation}
X_1=-i \left[ \tilde{p}_R^{(1)}+\tilde{p}_R^{(2)} \right]
G_{d,d}^{\rm r}(\omega)-ix_R e^{-i\phi} Y_1
\label{eq:X_1}
\end{equation}
with
\begin{equation}
Y_1 = \sum_{k}^{(1),(2)}\sqrt{w_{Lk}}G_{d,Lk}^{\rm r}(\omega).
\end{equation}
Using the Fourier transformation of Eq.\ (\ref{eq:Langreth1}),
we obtain
\begin{equation}
Y_1 = -i \left[ \tilde{p}_L^{(1)}+\tilde{p}_L^{(2)} \right]
G_{d,d}^{\rm r}(\omega)-ix_L e^{i\phi} X_1.
\label{eq:Y_1}
\end{equation}
From Eqs.\ (\ref{eq:X_1}) and (\ref{eq:Y_1}), we express
$X_1$ in terms of $G_{d,d}^{\rm r}(\omega)$.
In the same way, $X_2$ can be written using 
$G_{d,d}^{\rm r}(\omega)$ and $G_{d,d}^<(\omega)$.

A similar procedure is adopted for $Y_0$. The final result is
so lengthy that we show the current expression in the case of
Eq.\ (\ref{eq:mu_L_mu_R}), i.e.,
$\mu_{L}^{(1)}=\mu_{L}^{(2)} \equiv \mu_{L}$ and
$\mu_{R}^{(1)}=\mu_{R}^{(2)} \equiv \mu_{R}$.
After the variable conversion of $\hbar\omega \rightarrow
\varepsilon$,
\begin{eqnarray}
I_L^{(1)}=\frac{4e}{h} \int d\varepsilon
\Biggl\{
&-& \Gamma_L^{(1)} \left[
2f_L(\varepsilon) {\rm Im}G_{d,d}^{\rm r}(\varepsilon)
+{\rm Im}G_{d,d}^<(\varepsilon) \right]
+ x_L^{(1)}x_R \frac{2}{(1+x)^2}
\left[ f_L(\varepsilon) - f_R(\varepsilon) \right]
\nonumber \\
&+& \tilde{p}_L^{(1)} \left[ A_1 {\rm Re}G_{d,d}^{\rm r}(\varepsilon)
+A_2 {\rm Im}G_{d,d}^{\rm r}(\varepsilon)
+A_3 {\rm Im}G_{d,d}^<(\varepsilon) \right]
\nonumber \\
&+& x_L^{(1)} \left[ B_1 {\rm Re}G_{d,d}^{\rm r}(\varepsilon)
+B_2 {\rm Im}G_{d,d}^{\rm r}(\varepsilon)
+B_3 {\rm Im}G_{d,d}^<(\varepsilon) \right]
\Biggr\},
\label{eq:I_L_(1)_final}
\end{eqnarray}
where
\begin{eqnarray}
A_1 &=&
\frac{4}{(1+x)^2} \tilde{p}_R \cos\phi
\left[ f_L(\varepsilon) - f_R(\varepsilon) \right],
\\
A_2 &=&
\frac{4}{(1+x)^2} \left\{
f_L(\varepsilon)
\left[ x\tilde{p}_R \sin\phi + (2+x) x_R \tilde{p}_L \right]
+ f_R(\varepsilon)
\left[ \tilde{p}_R \sin\phi - x_R \tilde{p}_L \right]
\right\},
\\
A_3 &=&
\frac{2}{1+x}
\left( \tilde{p}_R \sin\phi + x_R \tilde{p}_L \right),
\\
B_1 &=&
-\frac{8}{(1+x)^3} x_R \tilde{p}_L \tilde{p}_R \cos\phi
\left[ f_L(\varepsilon) - f_R(\varepsilon) \right],
\\
B_2 &=&
\frac{2}{(1+x)^3} \Bigl\{
f_L(\varepsilon)
\left[ -2(1+x) x_R \tilde{p}_L \tilde{p}_R \sin\phi
+ (1-x) \tilde{p}_R^2 - (3+x) x_R^2 \tilde{p}_L^2 \right]
\nonumber \\
  & & -2 f_R(\varepsilon)
\left( \tilde{p}_R^2 - x_R^2 \tilde{p}_L^2 \right)
\Bigr\},
\\
B_3 &=&
-\frac{1}{(1+x)^2}
\left( 2 x_R \tilde{p}_L \tilde{p}_R \sin\phi + \tilde{p}_R^2
+x_R^2 \tilde{p}_L^2 \right),
\end{eqnarray}
with $\tilde{p}_{\alpha}
=\tilde{p}_{\alpha}^{(1)}+\tilde{p}_{\alpha}^{(2)}
=\sqrt{\Gamma_{\alpha} x_{\alpha}} p_{\alpha}$.

The current $I_L^{(2)}$ from lead $L(2)$ is given by
replacing $(1) \rightarrow (2)$ in Eq.\ (\ref{eq:I_L_(1)_final}).
The current $I_R^{(j)}$ from lead $R(j)$ is obtained from
$I_L^{(j)}$ by replacing $L \leftrightarrow R$
and $\phi \rightarrow -\phi$. These equations yield Eq.\
(\ref{eq:current_conserve}) for the current conservation.

\subsection{Current formula in terms of $G_{d,d}^{\rm r}$}

For the two-terminal model in Fig.\ \ref{Fig1:model0}(a),
the current from lead $L$ is $I_L=I_L^{(1)}+I_L^{(2)}$.
The elimination of $G_{d,d}^<$ using Eq.\ (\ref{eq:current_conserve})
results in its expression in Eq.\ (\ref{eq:2terminal_IL}).

As a three-terminal model, we examine the model in
Fig.\ \ref{Fig1:model0}(b) consisting
of leads $L$, $R(1)$, and $R(2)$. We introduce parameters,
$\gamma_{R}^{(j)}$, $y_{R}^{(j)}$, and $q_{R}^{(j)}$ in Eq.\
(\ref{eq:ratio_of_R(j)}).
The current into lead $R(1)$ is given by $-I_R^{(1)}$.
Eliminating $G_{d,d}^<$ using Eq.\ (\ref{eq:current_conserve}), we obtain
\begin{eqnarray}
-I_R^{(1)} &=& \frac{2e}{h} \int d\varepsilon
\left[ f_L(\varepsilon)-f_R(\varepsilon) \right]
T^{(1)}(\varepsilon) d\varepsilon
\label{eq:3terminal_IRa},
\\
T^{(1)}(\varepsilon) &=&
\frac{4x}{(1+x)^2}y_{R}^{(1)}
\nonumber \\
& &
+8\frac{(1+x)q_{R}^{(1)}-2xy_{R}^{(1)}}{(1+x)^3}
\sqrt{\Gamma_{L} \Gamma_{R}x} p_L p_R \cos \phi
{\rm Re} G_{d,d}^{\rm r}(\varepsilon)
\nonumber \\
& &
+\frac{4 C_2}{(1+x)^3\tilde{\Gamma}}
{\rm Im} G_{d,d}^{\rm r}(\varepsilon),
\label{eq:3terminal_IRb}
\end{eqnarray}
where
\begin{eqnarray}
C_2 &=&
-2(1+x)\sqrt{\Gamma_{L} \Gamma_{R}x} p_L p_R \sin \phi
\left[ x(q_{R}^{(1)}-y_{R}^{(1)}) \Gamma_L(1-p_L^2)
+(\gamma_{R}^{(1)}-q_{R}^{(1)})\Gamma_R \right]
\nonumber \\
& & 
+\frac{x^3}{1+x} y_{R}^{(1)}
\left[ (\Gamma_{L} p_L^2)^2 + (\Gamma_{R} p_R^2)^2 \right]
\nonumber \\
& & + x(1-x)y_{R}^{(1)}(\Gamma_{L} p_L)^2
+x \left[ (1+x)(-\gamma_{R}^{(1)}+2q_{R}^{(1)})-2xy_{R}^{(1)} \right]
 (\Gamma_{R} p_R)^2
\nonumber \\
& & -\Gamma_{L} \Gamma_{R} D_2,
\end{eqnarray}
with
\begin{eqnarray}
D_2 &=&
(1+x)^3 \gamma_{R}^{(1)} 
+4x\frac{(1+x)q_{R}^{(1)}-xy_{R}^{(1)}}{1+x} (p_L p_R)^2 \sin^2 \phi
\nonumber \\
& & +\frac{x^2[2(x+3)(x+1)q_{R}^{(1)}-(x^2+4x-3)y_{R}^{(1)}]}{1+x}
(p_L p_R)^2
\nonumber \\
& &  - x \left[ (x+1)(x+2) \gamma_{R}^{(1)}+2 y_{R}^{(1)} \right]
p_L^2
\nonumber \\
& & -x \left[ 2(x+1)(x+2) q_{R}^{(1)} -x(x+3) y_{R}^{(1)} \right]
p_R^2.
\end{eqnarray}

Regarding the $\phi$-dependence of the conductance at $T=0$,
Eqs.\ (\ref{eq:3terminal_IRa}) and (\ref{eq:3terminal_IRb}) yield
Eqs.\ (\ref{eq:G(1)_U0a}) and (\ref{eq:G(1)_U0b})
in the absence of $U$. In the presence of $U$, however, 
we cannot obtain such a simple form in general.

\section{Green's function in the presence of $U$}

For our models shown in Figs.\ \ref{Fig1:model0}(a) and (b),
the Green's function of the QD is solvable in the case of $U=0$.
As discussed in section II.D, the retarded Green's function is given by
\begin{eqnarray}
G_{d,d}^{\rm r}(\varepsilon)
= \frac{1}{\varepsilon - \tilde{\varepsilon}_{d}(\phi) +i \tilde{\Gamma}}
\label{eq:AppendB:Green0}
\end{eqnarray}
with the effective energy level $\tilde{\varepsilon}_{d}(\phi)$ in
Eq.\ (\ref{eq:effective_e_d}) and effective linewidth $\tilde{\Gamma}$
in Eq.\ (\ref{eq:effective_Gamma}).
The renormalization due to the direct tunneling between the leads and
the Aharonov-Bohm effect by the magnetic flux is included in
these effective parameters.

In the presence of $U$, we formulate the perturbation with respect to
the electron-electron interaction in the QD,
$H_U=U n_{\uparrow} n_{\downarrow}$. The Hamiltonian in Eq.\
(\ref{eq:Hamiltonian}) is divided into the non-interacting part $H_0$
and $H_U$; $H=H_0+H_U$. The contour-ordered Green's function of the QD,
$G_{d,d}^{\rm C}(\tau,\tau')=\langle \mathcal{T}_{\rm C}
d_{\sigma}(\tau) d_{\sigma}^{\dagger}(\tau')  \rangle/(i\hbar)$,
is written as
\begin{equation}
G_{d,d}^{\rm C} (\tau,\tau')
= \frac{1}{i \hbar} {\rm tr} \left\{ \rho_0
\mathcal{T}_{\rm C} \, d_{{\rm I},\sigma}(\tau)
d_{{\rm I},\sigma}^{\dagger}(\tau')
\exp\left[ \int_{\rm C}d\tau'' H_{{\rm I},U}(\tau'') \right] 
\right\},
\end{equation}
where $\rho_0$ is the density matrix for $U=0$ and index I indicates
the operator in the interaction picture,
$\mathcal{O}_{\rm I}(\tau)=e^{iH_0\tau/\hbar}\mathcal{O}e^{-iH_0\tau/\hbar}$.
In the perturbative expansion, the unperturbed Green's function is given
by Eq.\ (\ref{eq:AppendB:Green0}). This problem is equivalent to that of
the conventional Anderson impurity model, in which an impurity with
energy level $\tilde{\varepsilon}_{d}(\phi)$ and Coulomb interaction $U$
is connected to an energy-band of conduction electrons via
the effective hybridization $\tilde{\Gamma}$:
\begin{eqnarray}
H_{\rm Anderson}=
\tilde{\varepsilon}_{d}(\phi) \sum_{\sigma} n_{\sigma}
+ U n_{\uparrow} n_{\downarrow}
+ \sum_{k \sigma} \varepsilon_k a_{k,\sigma}^{\dagger}a_{k,\sigma}
+ \sum_{k \sigma} (v a_{k, \sigma}^{\dagger} d_{\sigma}+{\rm H.c.}),
\label{eq:AppendB:AndersonModel}
\end{eqnarray}
where $\tilde{\Gamma}=\pi \rho |v|^2$, with the density of states $\rho$
for the conduction electrons.

In the equilibrium with $eV=0$,
the physical quantities of electrons in our model can be evaluated
by exploiting the established methods for the Anderson impurity model
\cite{PhysRevLett.87.156803}. The retarded Green's function is given by
\begin{eqnarray}
G_{d,d}^{\rm r}(\varepsilon)
= \frac{1}{\varepsilon - \tilde{\varepsilon}_{d}(\phi) +i \tilde{\Gamma}
 - \Sigma_U(\varepsilon)}
\end{eqnarray}
with use of the self-energy $\Sigma_U(\varepsilon)$ due to the
electron-electron interaction in the QD.
Note that $z=[1-\frac{d\Sigma_U}{d\varepsilon}(0)]^{-1}$ and
$\tilde{\varepsilon}_d^*=z[\tilde{\varepsilon}_d(\phi) +\Sigma_U(0)]$
in Eq.\ (\ref{eq:Greenfunction_with_U}).
$G_{d,d}^{\rm r}(0)$ is expressed in
Eq.\ (\ref{eq:Greenfunction_sum_rule}) using the phase shift $\theta_{\rm QD}$.
The Friedel sum rule connects the phase shift to the electron occupation
per spin in the QD, $\theta_{\rm QD}= \pi\langle n_{\sigma}\rangle$,
where
\begin{eqnarray}
\langle n_{\sigma} \rangle
= \frac{1}{2}
 - \frac{1}{\pi} \tan^{-1} \left(
\frac{\tilde{\varepsilon}_d(\phi) +\Sigma_U(0)}{\tilde{\Gamma}} \right) \, .
\end{eqnarray}
We use the Bethe ansatz exact solution to evaluate
$\langle n_{\sigma}\rangle$ \cite{doi:10.1143/JPSJ.52.1119,Wiegmann_1983}.

\section{Current in three-terminal model in Fig.\ \ref{Fig4:model_3term}(b)}

We apply the current formula in Eqs.\ (\ref{eq:3terminal_IRa})
and (\ref{eq:3terminal_IRb}) to the model in Fig.\ \ref{Fig4:model_3term}(b).

As mentioned in section IV.C, 
$V_{R,k'}=V_R \alpha_R$
and $\sqrt{w_{R,k'}}=\sqrt{w_R}\beta_R$ when state $k'$
belongs to lead $R(1)$ while $V_{R,k'}=V_R \beta_R$
and $\sqrt{w_{R,k'}}=-\sqrt{w_R}\alpha_R$ when state $k'$ belongs to
lead $R(2)$ in the tunnel Hamiltonian $H_T$.
This results in $\Gamma_R^{(1)}=\alpha_R^2 \Gamma_R$,
$\Gamma_R^{(2)}=\beta_R^2 \Gamma_R$, $x_R^{(1)}=\beta_R^2 x_R$,
and $x_R^{(2)}=\alpha_R^2 x_R$. We also find that
$p_R^{(1)}=1$, $p_R^{(2)}=-1$, and hence $p_R=p_R^{(1)}+p_R^{(2)}=0$.

From $p_R=0$, $\tilde{\varepsilon}_d(\phi)=\varepsilon_d$
in Eq.\ (\ref{eq:effective_e_d}), which is independent of
the AB phase $\phi$ for the magnetic flux.
The Green's function in the absence of $U$ becomes
\begin{equation}
G_{d,d}^{\rm r}(\varepsilon)=
\frac{1}{\varepsilon-\varepsilon_d+i \tilde{\Gamma}}
\end{equation}
with
\begin{equation}
\tilde{\Gamma}= \Gamma_L \left(1- \frac{x}{1+x} p_L^2 \right) +\Gamma_R.
\end{equation}
The substitution of  $\gamma_{R}^{(1)}=\alpha_R^2$,
$y_{R}^{(1)} =\beta_R^2$, and $q_{R}^{(1)}p_R=\alpha_R \beta_R$
($q_R^{(1)}=\infty$) into Eq.\ (\ref{eq:3terminal_IRb}) results in
\begin{equation}
T^{(1)}(\varepsilon)=
\frac{4x}{(1+x)^2}\beta_R^2
+\frac{8 \alpha_R \beta_R}{(1+x)^2}
\sqrt{\Gamma_{L} \Gamma_{R} x} p_L \cos \phi
{\rm Re} G_{d,d}^{\rm r}(\varepsilon)
+\frac{4 C'_2}{(1+x)^3\tilde{\Gamma}}
{\rm Im} G_{d,d}^{\rm r}(\varepsilon),
\label{eq:T(1)_Fig5(b)}
\end{equation}
where
\begin{eqnarray}
C'_2 & = &
-2(1+x)\sqrt{\Gamma_{L} \Gamma_{R} x} p_L \sin \phi
[x\Gamma_{L}(1-p_L^2)-\Gamma_{R}] \alpha_R \beta_R
\nonumber \\
& & + \frac{x^3}{1+x}\beta_R^2 (\Gamma_{L} p_L^2)^2 +
x(1-x) \beta_R^2 (\Gamma_{L} p_L)^2
\nonumber \\
& & -\Gamma_{L} \Gamma_{R}
\left\{
(1+x)^3\alpha_R^2
-x [(x+1)(x+2) \alpha_R^2 + 2 \beta_R^2] p_L^2
\right\}.
\end{eqnarray}

Since $\tilde{\varepsilon}_d(\phi)=\varepsilon_d$ in this model,
Eq.\ (\ref{eq:measured_phase}) exactly holds in the absence of $U$,
which leads to Eq.\ (\ref{eq:measured_phase_b}).
Besides, even in the presence of $U$, a relation between
$\phi_{\rm max}$ and $\theta_{\rm QD}$ is derived in the following.
The substitution of Eq.\ (\ref{eq:Greenfunction_with_U})
into Eq.\ (\ref{eq:T(1)_Fig5(b)}) yields
\begin{equation}
T^{(1)}(0)=
\frac{8 \alpha_R \beta_R}{(1+x)^2} \sqrt{\Gamma_{L} \Gamma_{R} x} p_L
\frac{\tilde{\Gamma}^*}{\tilde{\Gamma}}
\frac{1}{(\tilde{\varepsilon}_d^*)^2+(\tilde{\Gamma}^*)^2} F_1(\phi)
+(\mbox{$\phi$-indep.\ terms})
\end{equation}
at $\varepsilon=E_{\rm F}=0$, where
\begin{equation}
F_1(\phi)=-\tilde{\varepsilon}_d^* \cos \phi +
[x\Gamma_{L}(1-p_L^2)-\Gamma_{R}]
\frac{\tilde{\Gamma}^*}{\tilde{\Gamma}} \sin \phi.
\end{equation}
For $\phi=\phi_{\rm max}$ at which $F_1(\phi)$ is maximal,
\begin{equation}
\tan \phi_{\rm max}
= \frac{-x\Gamma_{L}(1-p_L^2)+\Gamma_{R}}{\tilde{\Gamma}}
\tan \theta_{\rm QD},
\end{equation}
where $\tan \theta_{\rm QD}=\tilde{\Gamma}^*/\tilde{\varepsilon}_d^*$.
$\theta_{\rm QD}$ satisfies the Friedel sum rule in the presence of $U$.

The current to lead $R(2)$, $-I_R^{(2)}$, is given by
replacing $(1) \rightarrow (2)$ in Eq.\ (\ref{eq:3terminal_IRa}).
$T_R^{(2)}$ is obtained from $T_R^{(1)}$ in Eq.\ (\ref{eq:T(1)_Fig5(b)}),
replacing $\alpha_R \rightarrow \beta_R$ and
$\beta_R \rightarrow -\alpha_R$.
In $T_R^{(1)}$ and $T_R^{(2)}$, coefficients of $\cos\phi$ and $\sin\phi$
are the same in magnitude and opposite in sign.
As a result, the total current to leads $R(1)$ and $R(2)$ does not depend on
the AB phase $\phi$ for the magnetic flux:
\begin{equation}
-I_R^{(1)}-I_R^{(2)} =
\frac{2e}{h} \int
\left[ f_L(\varepsilon)-f_R(\varepsilon) \right] T(\varepsilon)
d\varepsilon,
\end{equation}
where
\begin{equation}
T(\varepsilon) =
T_R^{(1)}+T_R^{(2)}
=
\frac{4x}{(1+x)^2}+\frac{4 C_1}{(1+x)^3\tilde{\Gamma}}
{\rm Im} G_{d,d}^{\rm r}(\varepsilon),
\end{equation}
with
\begin{eqnarray}
C_1 &=&
\frac{x^3}{1+x} (\Gamma_{L} p_L^2)^2
+x(1-x) (\Gamma_{L} p_L)^2
\nonumber \\
& & 
-\Gamma_{L} \Gamma_{R} \left[ (1+x)^3 -x(x^2+3x+4) p_L^2 \right].
\end{eqnarray}
This coincides with Eq.\ (\ref{eq:2terminal_IL}) for the current in
the two-terminal system with $p_R=0$.

\section{Tight-binding model in Fig.\ \ref{Fig9:tight_binding}}

In the tight-binding model in Fig.\ \ref{Fig9:tight_binding}(a),
leads $L$ and $R$ consist of two sites in width and $N$ sites in length
($N \gg 1$). There are two subbands in the leads, as depicted in Fig.\
\ref{Fig9:tight_binding}(b),
\begin{equation}
\varepsilon_{\pm}(q)=\mp t_1 -2t \cos qa,
\end{equation}
where $t$ ($t_1$) is the transfer integral in $x$ ($y$) direction
and $a$ is the lattice constant.
$q$ is the wavenumber in the $x$ direction,
$q=\pi n/[(N+1)a]$ with $n=1,2,\cdots,N$. The corresponding
states are given by Eqs.\ (\ref{eq:TightBindingModel1}) and
(\ref{eq:TightBindingModel2}).

Let us consider the case of two conduction channels in the leads
when the Fermi level intersects both the two subbands. They
are labeled by $k=(q,\pm)$.
In the tunnel Hamiltonian $H_T$ in Eq.\ (\ref{eq:tunnelH}),
$V_{L;q,\pm}=V_L \psi_{L; q,\pm}(-1,1)$,
$V_{R;q,\pm}=V_R \psi_{R; q,\pm}(1,1)$,
and $W_{q',\gamma'; q,\gamma}=
\sqrt{w_{R; q',\gamma'} w_{L; q,\gamma}}$, where
$w_{L; q,\gamma}$ and $w_{R; q',\gamma'}$ are
given by Eqs.\ (\ref{eq:TightBindingModel3})
and (\ref{eq:TightBindingModel4}), respectively,
for $\gamma$, $\gamma'=\pm$.
Here,
$\psi_{\alpha; q,\gamma}(j,\ell)=
\langle j,\ell | \alpha;q,\gamma \rangle$ is 
the wavefunction of the conduction mode
$(q,\gamma)$ in lead $\alpha$:
$\psi_{L; q,\pm}(-1,1)=\psi_{R; q,\pm}(1,1)
=\sin qa/\sqrt{N+1}$ and 
$\psi_{L; q,\pm}(-1,2)=\psi_{R; q,\pm}(1,2)
=\pm \sin qa/\sqrt{N+1}$ from
Eqs.\ (\ref{eq:TightBindingModel1})
and (\ref{eq:TightBindingModel2}).

We calculate $\Gamma_{\alpha}$, $x_{\alpha}$, and
$p_{\alpha}$ in Eqs.\ (\ref{eq:Gamma})--(\ref{eq:introduce_p})
at $\varepsilon=E_{\rm F}$. We focus on lead $L$
because lead $R$ is identical to lead $L$.
The density of states for subband $\pm$ is given by
\begin{equation}
\rho_{\pm}(E_{\rm F})=\frac{N+1}{\pi}\frac{1}{2t\sin q_{\pm}a},
\end{equation}
where $q_{\pm}$ is defined by
$\varepsilon_{\pm}(q_{\pm})=E_{\rm F}$, as depicted in Fig.\
\ref{Fig9:tight_binding}(b),
\begin{eqnarray}
\Gamma_L &=&
\pi \sum_{\gamma=\pm} \rho_{\gamma}(E_{\rm F})
(V_{L;q_{\gamma},\gamma})^2
\nonumber
\\
&=& \frac{(V_L)^2}{2t}(\sin q_+a + \sin q_-a),
\label{eq:AppendixC1}
\\
x_L
&=& \frac{W}{2t}(\sin q_+a + \sin q_-a),
\\
\sqrt{\Gamma_L x_L} p_L
&=& \frac{V_L\sqrt{W}}{2t}(\sin q_+a - \sin q_-a),
\label{eq:AppendixC2}
\end{eqnarray}
and in consequence we obtain $p_L$ in Eq.\ (\ref{eq:TightBindingModel5}).

\end{document}